%% file: main.tex
\pdfoutput=1
\documentclass[12pt,a4paper]{article}

\usepackage{ifthen} 
\newboolean{pdflatex}
\setboolean{pdflatex}{true} 

\newboolean{articletitles}
\setboolean{articletitles}{true} 

\newboolean{uprightparticles}
\setboolean{uprightparticles}{false} 

\def\paperauthors{LHCb collaboration} 
\def\paperasciititle{Amplitude analysis of B0 to  eta_c(1S) K+ pi- decays} 
\def\papertitle{Amplitude analysis of $B^0 \rightarrow \eta_c(1S) K^+ \pi^- $ decays} 
\def\paperkeywords{{High Energy Physics}, {LHCb}} 
\def\papercopyright{\the\year\ CERN for the benefit of the LHCb collaboration} 
\def\paperlicence{CC BY 4.0 licence}
\def\paperlicenceurl{https://creativecommons.org/licenses/by/4.0/}

\newif\ifEnableSectionTOCLinks
\EnableSectionTOCLinksfalse 

\input{preamble}
\usepackage{longtable} 
\usepackage{rotating}

\begin{document}

\renewcommand{\thefootnote}{\fnsymbol{footnote}}
\setcounter{footnote}{1}

\input{title-LHCb-PAPER}


\renewcommand{\thefootnote}{\arabic{footnote}}
\setcounter{footnote}{0}


\cleardoublepage


\pagestyle{plain} 
\setcounter{page}{1}
\pagenumbering{arabic}


\input{body}

\input{acknowledgements}




\addcontentsline{toc}{section}{References}
\bibliographystyle{LHCb}
\bibliography{main,standard,LHCb-PAPER,LHCb-CONF,LHCb-DP,LHCb-TDR}

\newpage
\input{Authorship_LHCb-PAPER-2025-027}

\end{document}

%% file: preamble.tex
\usepackage[top=1in, bottom=1.25in, left=1in, right=1in]{geometry}

\usepackage{microtype}
\usepackage{lineno}  
\usepackage{xspace} 
\usepackage{caption} 

\usepackage{graphicx}  
\usepackage{color}
\usepackage{colortbl}
\graphicspath{{./figs/}} 

\usepackage{amsmath} 
\usepackage{amssymb}
\usepackage{amsfonts}
\usepackage{upgreek} 

\newcommand*\patchAmsMathEnvironmentForLineno[1]{%
\expandafter\let\csname old#1\expandafter\endcsname\csname #1\endcsname
\expandafter\let\csname oldend#1\expandafter\endcsname\csname
end#1\endcsname
 \renewenvironment{#1}%
   {\linenomath\csname old#1\endcsname}%
   {\csname oldend#1\endcsname\endlinenomath}%
}
\newcommand*\patchBothAmsMathEnvironmentsForLineno[1]{%
  \patchAmsMathEnvironmentForLineno{#1}%
  \patchAmsMathEnvironmentForLineno{#1*}%
}
\AtBeginDocument{%
\patchBothAmsMathEnvironmentsForLineno{equation}%
\patchBothAmsMathEnvironmentsForLineno{align}%
\patchBothAmsMathEnvironmentsForLineno{flalign}%
\patchBothAmsMathEnvironmentsForLineno{alignat}%
\patchBothAmsMathEnvironmentsForLineno{gather}%
\patchBothAmsMathEnvironmentsForLineno{multline}%
\patchBothAmsMathEnvironmentsForLineno{eqnarray}%
}

\usepackage[pdftex,
            pdfauthor={\paperauthors},
            pdftitle={\paperasciititle},
            pdfkeywords={\paperkeywords}]{hyperref}
\usepackage{hyperxmp}
\hypersetup{
    pdfcopyright={Copyright (C) \papercopyright},
    pdflicenseurl={\paperlicenceurl}
}

\usepackage[colorinlistoftodos,textsize=scriptsize]{todonotes}

\usepackage[bottom,flushmargin,hang,multiple]{footmisc}

\usepackage[all]{hypcap} 

\input{lhcb-symbols-def} 

\hypersetup{
  colorlinks   = true, 
  urlcolor     = blue, 
  linkcolor    = blue, 
  citecolor    = red   
}

\ifEnableSectionTOCLinks
    \usepackage[explicit]{titlesec} 
    
    \let\oldcontentsline\contentsline
    \renewcommand

    \titleformat{\section}{\normalfont\Large\bf}{\hyperlink{tocsection.\thesection}{{\thesection} \parbox[t]{\dimexpr\textwidth-1pc}{#1}}}{1pc}{}

    \titleformat{\subsection}{\normalfont\bf}{\hyperlink{tocsubsection.\thesubsection}{{\thesubsection} \parbox[t]{\dimexpr\textwidth-1pc}{#1}}}{1pc}{}

    \titleformat{name=\section,numberless}[display]{}{}{0pt}{\normalfont\Huge\bfseries #1}
\fi

\usepackage{cite} 
\usepackage{mciteplus}

%% file: lhcb-symbols-def.tex
\usepackage{xspace}
\usepackage{upgreek}

\def\lhcb   {\mbox{LHCb}\xspace}

\def\belle  {\mbox{Belle}\xspace}

\def\besiii {\mbox{BESIII}\xspace}
\def\cleo   {\mbox{CLEO}\xspace}

\def\MagUp {\mbox{\em Mag\kern -0.05em Up}\xspace}

\ifthenelse{\boolean{uprightparticles}}%
{

 \def\Peta        {\ensuremath{\upeta}\xspace}

 \def\Ppi         {\ensuremath{\uppi}\xspace}

 \def\Pphi        {\ensuremath{\upphi}\xspace}
 
 \def\Pchi        {\ensuremath{\upchi}\xspace}
 \def\Ppsi        {\ensuremath{\uppsi}\xspace}

 \def\PDelta      {\ensuremath{\Delta}\xspace}
 \def\PXi         {\ensuremath{\Xi}\xspace}
 \def\PLambda     {\ensuremath{\Lambda}\xspace}
 \def\PSigma      {\ensuremath{\Sigma}\xspace}
 \def\POmega      {\ensuremath{\Omega}\xspace}
 \def\PUpsilon    {\ensuremath{\Upsilon}\xspace}
 \let\oldPi\Pi
 \def\PPi         {\ensuremath{\oldPi}\xspace}

 \def\PB      {\ensuremath{\mathrm{B}}\xspace}
 \def\PD      {\ensuremath{\mathrm{D}}\xspace}
 \def\PJ      {\ensuremath{\mathrm{J}}\xspace}
 \def\PK      {\ensuremath{\mathrm{K}}\xspace}
 \def\Pb      {\ensuremath{\mathrm{b}}\xspace}
 \def\Pc      {\ensuremath{\mathrm{c}}\xspace}

 \def\Pp      {\ensuremath{\mathrm{p}}\xspace}
 \def\Pq      {\ensuremath{\mathrm{q}}\xspace}
 \def\Ps      {\ensuremath{\mathrm{s}}\xspace}

 \def\thebaroffset{0.0em}
}
{

 \def\Peta        {\ensuremath{\eta}\xspace}

 \def\Ppi         {\ensuremath{\pi}\xspace}

 \def\Pphi        {\ensuremath{\phi}\xspace}
 
 \def\Pchi        {\ensuremath{\chi}\xspace}
 \def\Ppsi        {\ensuremath{\psi}\xspace}
 
 \mathchardef\PDelta="7101
 \mathchardef\PXi="7104
 \mathchardef\PLambda="7103
 \mathchardef\PSigma="7106
 \mathchardef\POmega="710A
 \mathchardef\PUpsilon="7107
 \mathchardef\PPi="7105
 \def\PB      {\ensuremath{B}\xspace}
 \def\PD      {\ensuremath{D}\xspace}
 \def\PJ      {\ensuremath{J}\xspace}
 \def\PK      {\ensuremath{K}\xspace}
 \def\Pb      {\ensuremath{b}\xspace}
 \def\Pc      {\ensuremath{c}\xspace}

 \def\Pp      {\ensuremath{p}\xspace}
 \def\Pq      {\ensuremath{q}\xspace}
 \def\Ps      {\ensuremath{s}\xspace}

 \def\thebaroffset{0.18em}
}
\newcommand{\offsetoverline}[2][\thebaroffset]{\kern #1\overline{\kern -#1 #2}}%

\makeatletter
\ifcase \@ptsize \relax
  \newcommand{\miniscule}{\@setfontsize\miniscule{4}{5}}
\or
  \newcommand{\miniscule}{\@setfontsize\miniscule{5}{6}}
\or
  \newcommand{\miniscule}{\@setfontsize\miniscule{5}{6}}
\fi
\makeatother

\DeclareRobustCommand{\optbar}[1]{\shortstack{{\miniscule (\rule[.5ex]{1.25em}{.18mm})}
  \\ [-.7ex] $#1$}}

\def\quark     {{\ensuremath{\Pq}}\xspace}
\def\quarkbar  {{\ensuremath{\overline \quark}}\xspace}

\def\squark    {{\ensuremath{\Ps}}\xspace}

\def\cquark    {{\ensuremath{\Pc}}\xspace}
\def\cquarkbar {{\ensuremath{\overline \cquark}}\xspace}

\def\bquark    {{\ensuremath{\Pb}}\xspace}

\def\pion   {{\ensuremath{\Ppi}}\xspace}

\def\pip    {{\ensuremath{\pion^+}}\xspace}
\def\pim    {{\ensuremath{\pion^-}}\xspace}

\def\kaon    {{\ensuremath{\PK}}\xspace}

\def\KorKbar {\kern \thebaroffset\optbar{\kern -\thebaroffset \PK}{}\xspace}

\def\Kp      {{\ensuremath{\kaon^+}}\xspace}

\def\Kstarz  {{\ensuremath{\kaon^{*0}}}\xspace}

\def\Kstar   {{\ensuremath{\kaon^*}}\xspace}

\newcommand{\phiz}{\ensuremath{\Pphi}\xspace}

\def\Dbar    {{\ensuremath{\offsetoverline{\PD}}}\xspace}
\def\D       {{\ensuremath{\PD}}\xspace}

\def\DorDbar {\kern \thebaroffset\optbar{\kern -\thebaroffset \PD}\xspace}

\def\Dzb     {{\ensuremath{\Dbar{}^0}}\xspace}
\def\Dp      {{\ensuremath{\D^+}}\xspace}
\def\Dm      {{\ensuremath{\D^-}}\xspace}

\def\DpDm    {\ensuremath{\Dp {\kern -0.16em \Dm}}\xspace}
\def\Dstar   {{\ensuremath{\D^*}}\xspace}
\def\Dstarb  {{\ensuremath{\Dbar{}^*}}\xspace}

\def\B       {{\ensuremath{\PB}}\xspace}

\def\BorBbar {\kern \thebaroffset\optbar{\kern -\thebaroffset \PB}\xspace}
\def\Bz      {{\ensuremath{\B^0}}\xspace}

\def\Bd      {{\ensuremath{\B^0}}\xspace}

\def\BdorBdbar {\kern \thebaroffset\optbar{\kern -\thebaroffset \Bd}\xspace}

\def\Bs      {{\ensuremath{\B^0_\squark}}\xspace}

\def\BsorBsbar {\kern \thebaroffset\optbar{\kern -\thebaroffset \Bs}\xspace}

\def\jpsi     {{\ensuremath{{\PJ\mskip -3mu/\mskip -2mu\Ppsi}}}\xspace}

\def\etac     {{\ensuremath{\Peta_\cquark}}\xspace}

\def\chicone  {{\ensuremath{\Pchi_{\cquark 1}}}\xspace}

\def\Y#1S{\ensuremath{\PUpsilon{(#1S)}}\xspace}

\def\proton      {{\ensuremath{\Pp}}\xspace}
\def\antiproton  {{\ensuremath{\overline \proton}}\xspace}

\def\Lbar        {{\ensuremath{\offsetoverline{\PLambda}}}\xspace}
\def\LorLbar     {\kern \thebaroffset\optbar{\kern -\thebaroffset \PLambda}\xspace}

\def\Lcbar       {{\ensuremath{\Lbar{}^-_\cquark}}\xspace}

\def\BF         {{\ensuremath{\mathcal{B}}}\xspace}

\newcommand{\decay}[2]{\mbox{\ensuremath{#1\!\to #2}}\xspace}

\def\to                 {\ensuremath{\rightarrow}\xspace}

\def\AT#1     {\ensuremath{A_{\mathrm{T}}^{#1}}\xspace}           

\def\C#1      {\ensuremath{\mathcal{C}_{#1}}\xspace}                       
\def\Cp#1     {\ensuremath{\mathcal{C}_{#1}^{'}}\xspace}                    
\def\Ceff#1   {\ensuremath{\mathcal{C}_{#1}^{\mathrm{(eff)}}}\xspace}        
\def\Cpeff#1  {\ensuremath{\mathcal{C}_{#1}^{'\mathrm{(eff)}}}\xspace}       
\def\Ope#1    {\ensuremath{\mathcal{O}_{#1}}\xspace}                       
\def\Opep#1   {\ensuremath{\mathcal{O}_{#1}^{'}}\xspace}                    

\newcommand{\nospaceunit}[1]{\ensuremath{\text{#1}}}
\newcommand{\aunit}[1]{\ensuremath{\text{\,#1}}}

\newcommand{\tev}{\aunit{Te\kern -0.1em V}\xspace}
\newcommand{\gev}{\aunit{Ge\kern -0.1em V}\xspace}
\newcommand{\mev}{\aunit{Me\kern -0.1em V}\xspace}
\newcommand{\kev}{\aunit{ke\kern -0.1em V}\xspace}
\newcommand{\ev}{\aunit{e\kern -0.1em V}\xspace}

\newcommand{\mevc}{\ensuremath{\aunit{Me\kern -0.1em V\!/}c}\xspace}
\newcommand{\gevc}{\ensuremath{\aunit{Ge\kern -0.1em V\!/}c}\xspace}
\newcommand{\mevcc}{\ensuremath{\aunit{Me\kern -0.1em V\!/}c^2}\xspace}
\newcommand{\gevcc}{\ensuremath{\aunit{Ge\kern -0.1em V\!/}c^2}\xspace}

\def\mum  {\ensuremath{\,\upmu\nospaceunit{m}}\xspace}

\def\fb   {\ensuremath{\aunit{fb}}\xspace}
\def\invfb   {\ensuremath{\fb^{-1}}\xspace}

\newcommand{\chisq}{\ensuremath{\chi^2}\xspace}
\newcommand{\chisqndf}{\ensuremath{\chi^2/\mathrm{ndf}}\xspace}
\newcommand{\chisqip}{\ensuremath{\chi^2_{\text{IP}}}\xspace}

\def\gsim{{~\raise.15em\hbox{$>$}\kern-.85em
          \lower.35em\hbox{$\sim$}~}\xspace}
\def\lsim{{~\raise.15em\hbox{$<$}\kern-.85em
          \lower.35em\hbox{$\sim$}~}\xspace}

\def\sPlot{\mbox{\em sPlot}\xspace}

\def\sqs   {\ensuremath{\protect\sqrt{s}}\xspace}

\def\pt         {\ensuremath{p_{\mathrm{T}}}\xspace}

\def\ptot       {\ensuremath{p}\xspace}

\def\evtgen     {\mbox{\textsc{EvtGen}}\xspace}

\def\geant      {\mbox{\textsc{Geant4}}\xspace}

\def\photos     {\mbox{\textsc{Photos}}\xspace}

\def\pythia     {\mbox{\textsc{Pythia}}\xspace}

\def\roofit     {\mbox{\textsc{RooFit}}\xspace}

\def\tell1  {TELL1\xspace}
\def\ukl1   {UKL1\xspace}

\newcommand{\ie}{\mbox{\itshape i.e.}\xspace}

\newcommand{\lhcborcid}[1]{\href{https://orcid.org/#1}{\hspace*{0.1em}\raisebox{-0.45ex}{\includegraphics[width=1em]{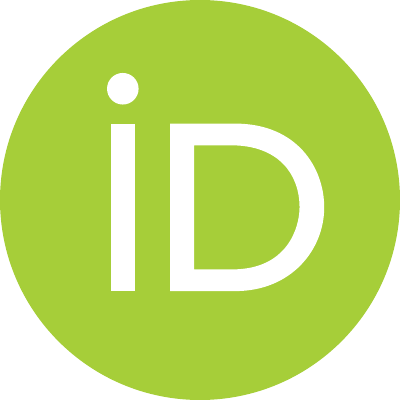}}}}


%% file: title-LHCb-PAPER.tex

\begin{titlepage}
\pagenumbering{roman}

\vspace*{-1.5cm}
\centerline{\large EUROPEAN ORGANIZATION FOR NUCLEAR RESEARCH (CERN)}
\vspace*{1.5cm}
\noindent
\begin{tabular*}{\linewidth}{lc@{\extracolsep{\fill}}r@{\extracolsep{0pt}}}
\ifthenelse{\boolean{pdflatex}}
{\vspace*{-1.5cm}\mbox{\!\!\!\includegraphics[width=.14\textwidth]{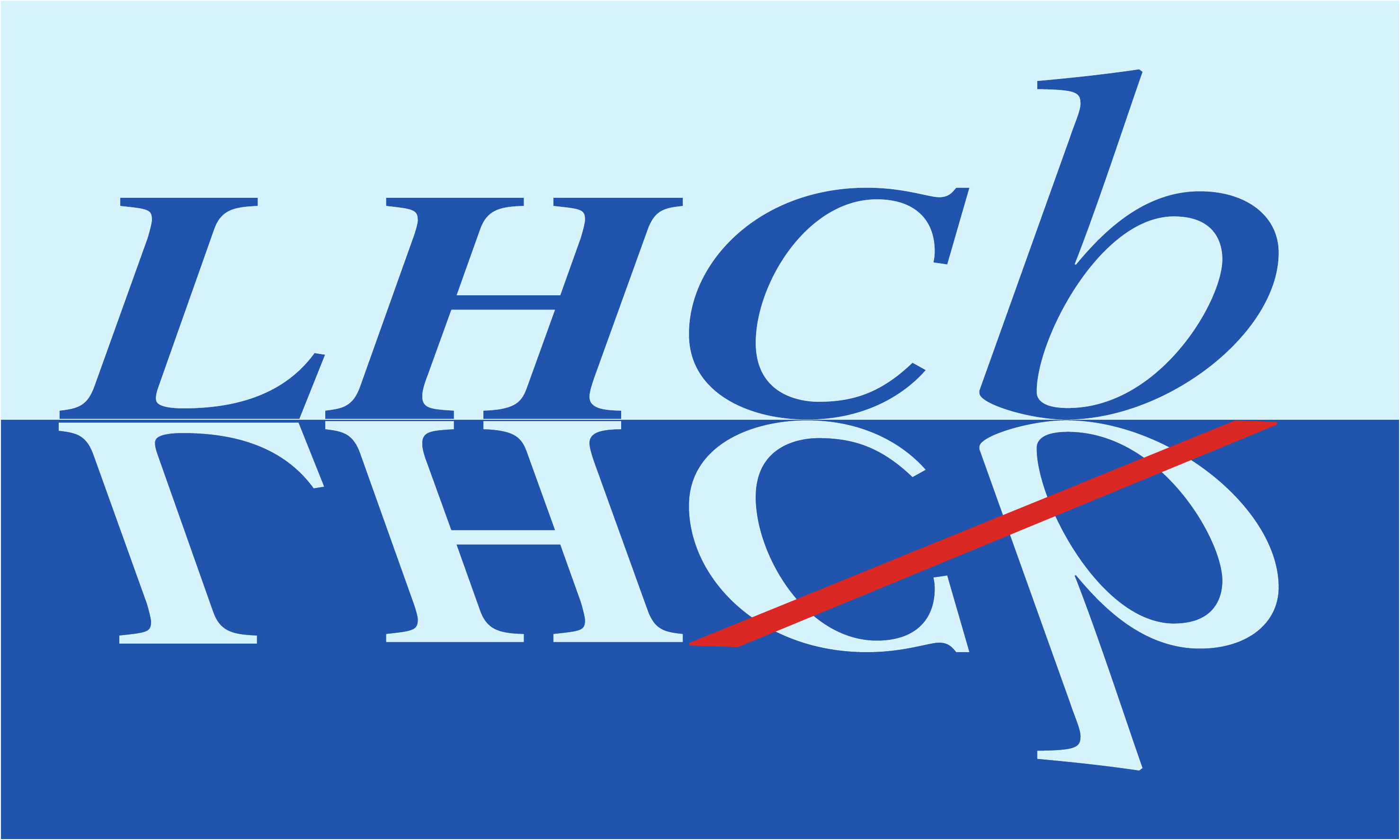}} & &}%
{\vspace*{-1.2cm}\mbox{\!\!\!\includegraphics[width=.12\textwidth]{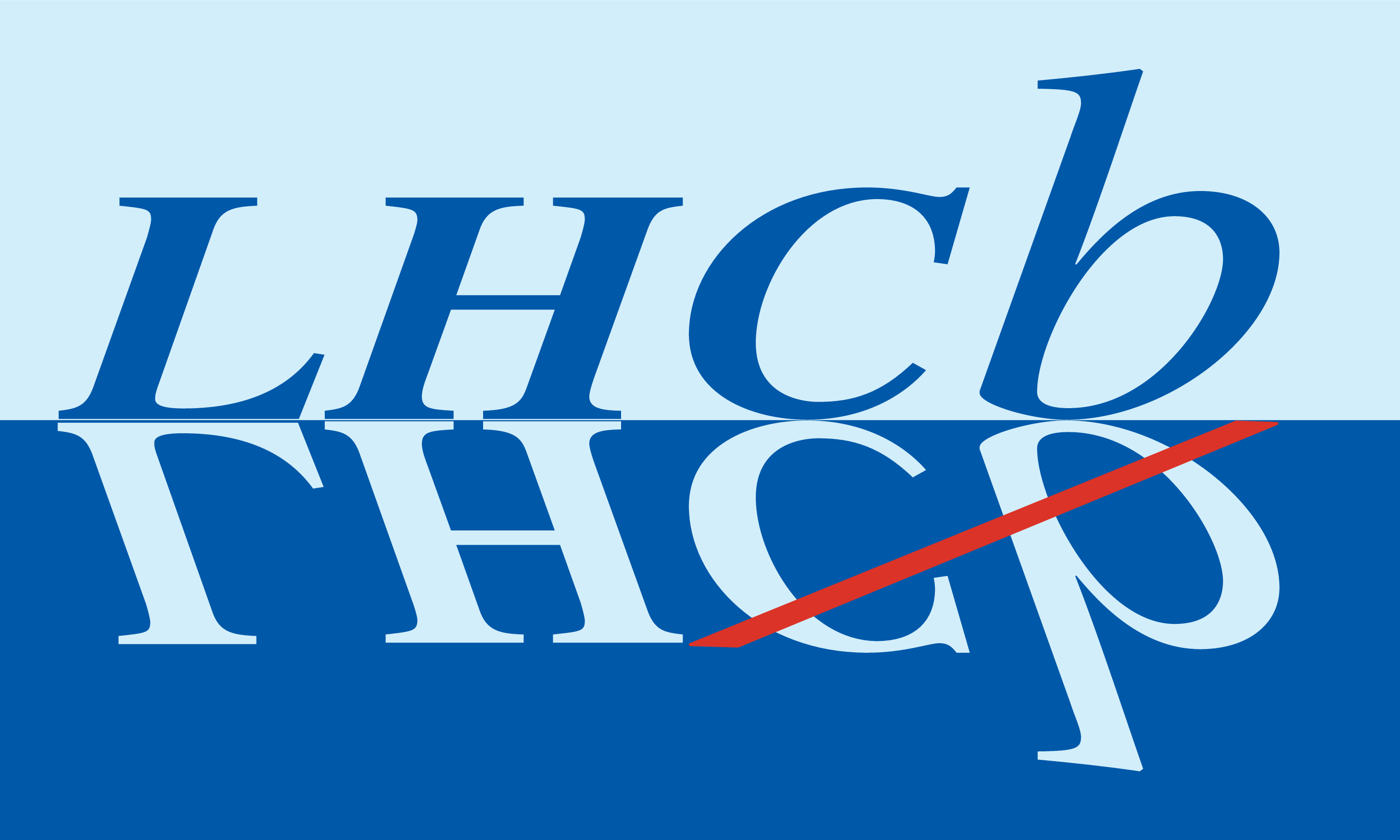}} & &}%
\\
 & & CERN-EP-2025-185 \\  
 & & LHCb-PAPER-2025-027 \\  
 & & September 3, 2025 \\ 
 & & \\
\end{tabular*}

\vspace*{4.0cm}

{\normalfont\bfseries\boldmath\huge
\begin{center}
  \papertitle 
\end{center}
}

\vspace*{2.0cm}

\begin{center}
\paperauthors\footnote{Authors are listed at the end of this paper.}
\end{center}

\vspace{\fill}

\begin{abstract}
  \noindent
An amplitude analysis of the  $\decay{\Bd}{ \etac(1S) \Kp \pim} $ decays with $\etac(1S) \to \proton \antiproton$ is performed using a sample corresponding to an integrated luminosity of 9\invfb of \proton\proton collision data collected by the LHCb detector at centre-of-mass energies of \mbox{\sqs = 7,} 8 and 13\tev. The data are described with a model including only intermediate contributions from known \Kstarz resonances. Evidence for an exotic resonance in the $\etac(1S) \pim $ system, reported in a previous analysis of this decay channel, is not confirmed. The inclusive branching fraction of the $\Bd \rightarrow \etac(1S) \Kp \pim $ decays is measured to be
\begin{align*}
\mathcal{B}(\Bd \to \etac(1S) \Kp \pim) = (5.82 \pm 0.20 \pm 0.23 \pm 0.55) \times 10^{-4},
\end{align*}
where the first uncertainty is statistical, the second systematic, and the third arises from the limited knowledge of external branching fractions.
\end{abstract}

\vspace*{2.0cm}

\begin{center}
  'Published in The European Physical Journal C Volume 86, article number 537, (2026)'
\end{center}

\vspace{\fill}

{\footnotesize 
\centerline{\copyright~\papercopyright. \href{\paperlicenceurl}{\paperlicence}.}}
\vspace*{2mm}

\end{titlepage}


\newpage
\setcounter{page}{2}
\mbox{~}
%
%
%
%

%% file: body.tex
\section{Introduction}

\label{sec:Introduction} 

Since the formulation of the quark model, hadronic states beyond the conventional $\quark\quarkbar$ mesons and \quark\quark\quark baryons have been proposed. 
Models based on quantum chromodynamics  predict the existence of various nonconventional hadrons, composed of different combinations of quarks and gluons, including glueballs, molecular states and tightly bound states, such as pentaquarks~(\quark\quark\quark\quark\quarkbar) and tetraquarks~(\quark\quark\quarkbar\quarkbar )~\cite{Gell-Mann:1964ewy,Zweig:352337}.
In the early 2000s, a hadron with unexpected properties, the $\chicone(3872)$ meson, was observed~\cite{Belle:2003nnu}, followed shortly by the discovery of many other charmonium-like and bottomonium-like states~\cite{Belle:2011aa,Belle:2004lle,CLEO:2006ike,CDF:2011pep}.
In some cases the properties of these states, including their electric charge, are inconsistent with predictions from conventional \quark\quarkbar models. Various interpretations have been proposed regarding their nature, binding and production mechanisms, and internal structure~\cite{Karliner:2015ina,Maiani:2004vq,Maiani:2015vwa,Dubynskiy:2008mq,Bugg:2011jr,Pakhlov:2014qva}.
To advance the understanding of these exotic hadrons, it is essential to investigate alternative production mechanisms and decay modes of previously observed unconventional states, while also searching for new candidates.

The \besiii collaboration observed in the $\jpsi \pi^+\pi^-$ final state a resonance whose minimal composition requires
the presence of two quarks and two antiquarks~\cite{BESIII:2013ris}, denoted $T_{\cquark\cquarkbar1}(3900)^{-}$, later confirmed also by the \belle~\cite{Belle:2013yex} and \cleo collaborations~\cite{Xiao:2013iha}.
This state has been interpreted as a hadrocharmonium state with \jpsi embedded in excited light-quark matter~\cite{Voloshin:2013dpa}, implying a partner state where the \jpsi is replaced by the $\eta_c(1S)$. Such a state would correspond to an isovector resonance decaying to $\etac(1S) \pim$ with similar structure and quantum numbers.\footnote{The simplified notation $\eta_c$ is used to refer to the $\eta_c(1S)$ meson.}
This interpretation is further supported by heavy-quark spin symmetry~\cite{Voloshin:2013dpa}, lattice QCD~\cite{Braaten:2013boa}, and diquark models~\cite{Maiani:2004vq}, which all foresee such a decay channel.

Evidence of a state decaying to $\etac \pim$, named $T_{\cquark\cquarkbar}(4100)^-$, has been found in the previous Dalitz Plot (DP)~\cite{Dalitz:1953cp} analysis of $\Bd \rightarrow \etac\Kp\pim $ decays by the LHCb collaboration~\cite{LHCb-PAPER-2018-034} using a data sample of \proton\proton collisions corresponding to an integrated luminosity of 4.7\invfb, collected in 2011, 2012 and 2016 at centre-of-mass energies of \sqs = 7, 8, and 13\tev, respectively.\footnote{The inclusion of charge-conjugate processes is implied
throughout.} 
The mass and natural width of this resonance are measured to be $4096 \pm  20^{+18}_{-22}\mev$ and $ 152  \pm  58^{+60}_{-35}\mev$, respectively.\footnote{Natural units with $\hbar = c = 1$ are used throughout.} 
Since this would make the state significantly more massive than the $T_{\cquark\cquarkbar1}(3900)^-$ state, it cannot be its expected partner.
Following the evidence reported by the LHCb collaboration for the $T_{c\bar{c}}(4100)^-$ charmonium-like candidate, several theoretical studies have been conducted to investigate its nature. The authors of Ref.~\cite{Sundu:2018nxt} calculated the mass, width, and coupling of this resonance, treating it as a scalar four-quark system with a diquark–antidiquark structure. Other studies found the $T_{c\bar{c}}(4100)^{-}$ state to be compatible with the tetraquark hypothesis favouring the quantum number $J^{PC}=0^{++}$ based on the measured width~\cite{Wu:2018xdi}. The study reported in Ref.~\cite{Voloshin:2018vym} suggests that the state can be described as hadrocharmonium, where the $\etac $ charmonium core is embedded in an S-wave configuration inside an excited light-quark state with the quantum numbers of a pion, $I^G(J^P)=1^-(0^-)$.
It further proposes that the mass difference between the $T_{c\bar{c}1}(4200)^{-}$, observed in the $\jpsi\Kp\pim$ final state~\cite{Belle:2014nuw,LHCb-PAPER-2018-043}, and $T_{c\bar{c}}(4100)^{-}$ states should be similar to that between the $ \jpsi$ and $\etac $ mesons, and that the partial widths of their dominant decays to $ \jpsi \pim$ and $\etac  \pim$, respectively, should be the same.
There are also studies~\cite{PhysRevD.100.054004} proposing that $T_{c\bar{c}}(4100)^{-}$ is the charge-conjugate state of the $T_{c\bar{c}}(4050)^{+}$ state observed by the Belle collaboration in the $\chi_{c1}\pip$ final state~\cite{ PhysRevD.78.072004}.
Finally, it has been suggested that the observed structure could be a kinematic effect arising from an S-wave \Dstar \Dstarb pair rescattering with $I^G(J^{PC}) = 1^-(0^{++})$ or a resonance produced by the P-wave \Dstar \Dstarb interaction~\cite{Zhao:2018xrd}.

\begin{figure}[tb]
\centering
\includegraphics[width=0.48\textwidth]{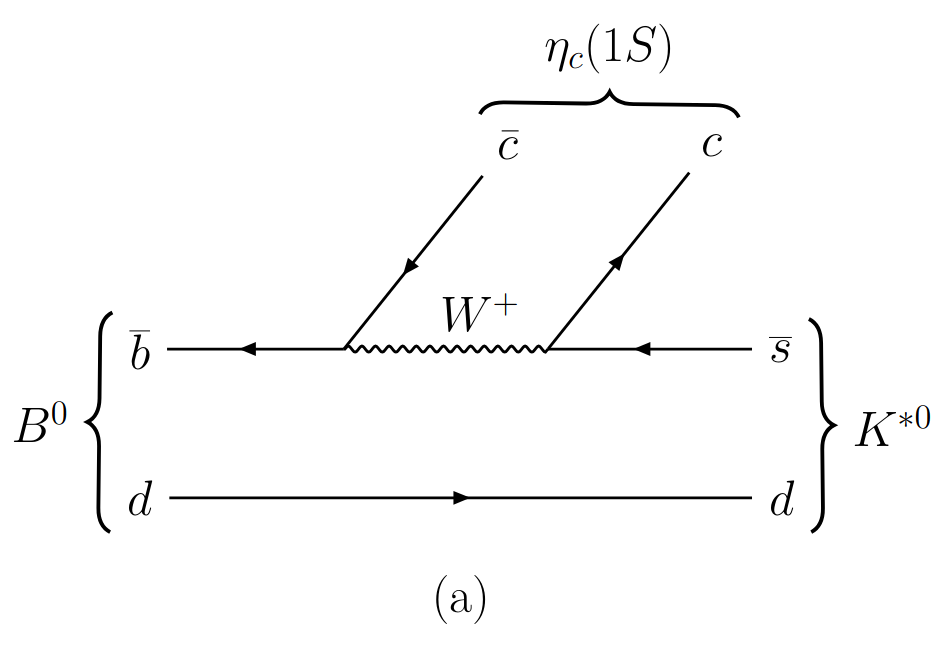}
\includegraphics[width=0.48\textwidth]{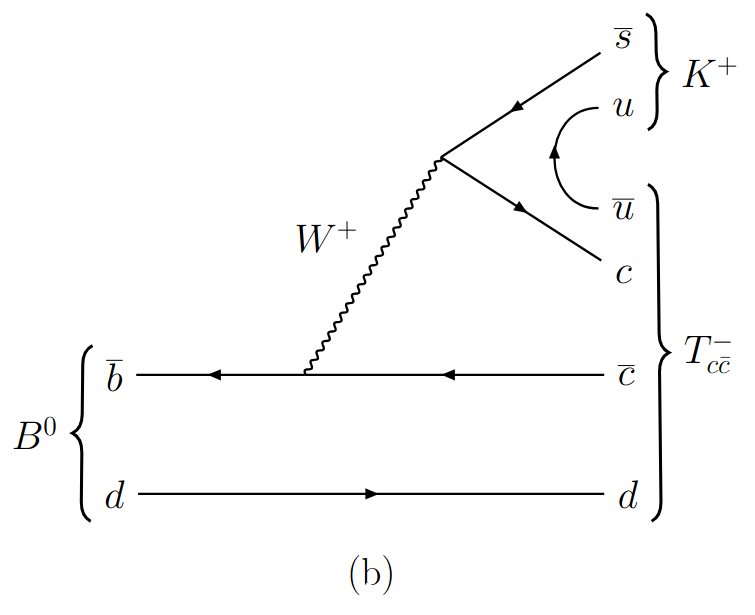}
\caption{Feynman diagrams for the (a) $\Bd  \to \etac(1S) K^{*0}$ and (b) $\Bd  \to T_{c\bar{c}}(4100)^{-}\Kp $ decay modes.}
\label{feymann}
\end{figure}

This letter reports an updated study of the $\Bd  \to \etac  \Kp  \pim $ decay with a dataset approximately twice the size of the one used in Ref.~\cite{LHCb-PAPER-2018-034} which enables a more precise evaluation of the presence of exotic contributions in this decay, as well as a more accurate determination of their properties.
The $B^0 \to \eta_c K^{+} \pi^{-}$ decay is expected to be dominated by the $K^*(892)^0 \to K^{+} \pi^{-}$ resonance, with some contributions from other $K^{*0}\to K^{+} \pi^{-}$ resonances. The decay could also proceed through exotic intermediate states, which could result in resonant structures in the $\eta_c\pi^{-}$ system. The Feynman diagrams of the $B^0 \to \eta_c K^{+} \pi^{-}$ decay are shown in Fig.~\ref{feymann}.
The $\Bd \to \etac \Kp \pim$ decay is studied by reconstructing the $\etac$ meson through its \proton\antiproton decay mode rather than the dominant mesonic modes.
This choice avoids the need to distinguish  between kaons and pions originating from $\Bd$ decay, reducing associated systematic uncertainties~\cite{Belle:2019avj}.
The $\Bd  \to \etac  \Kp  \pim$ decay involves only pseudoscalar mesons in the initial and final states and it is therefore fully described by two independent kinematic variables. This allows for a  DP analysis, which provides a complete characterisation of the decay dynamics in the assumption of no sizeable effects due to additional interfering amplitudes, involving suppressed Flavour Changing Neutral Currents, with the same $\Bd  \to \proton \antiproton  \Kp  \pim$ final state.
The absolute branching fraction of the $\Bd \to \eta_c\Kp\pim$ decay is also measured, using the $\Bd \to \jpsi\Kp\pim$ decay as normalisation channel.
The data sample corresponds to an integrated luminosity of 9\invfb of \proton\proton collision data collected by the LHCb detector at centre-of-mass energies of \sqs = 7, 8, and 13\tev in 2011, 2012 and 2015--2018, respectively. Data collected in 2011 and 2012 are referred to as Run~1 data, while data collected in 2015--2018 are referred to as Run~2 data.

\section{Detector and simulation}
\label{DetectorSimulation}

The \lhcb detector~\cite{LHCb-DP-2008-001,LHCb-DP-2014-002} is a single-arm forward
spectrometer covering the \mbox{pseudorapidity} range $2<\eta <5$,
designed for the study of particles containing \bquark or \cquark
quarks. The detector used to collect the data analysed in this paper includes a high-precision tracking system
consisting of a silicon-strip vertex detector surrounding the $pp$
interaction region~\cite{LHCb-DP-2014-001}, a large-area silicon-strip detector located
upstream of a dipole magnet with a bending power of about
$4{\mathrm{\,T\,m}}$, and three stations of silicon-strip detectors and straw
drift tubes~\cite{LHCb-DP-2013-003,LHCb-DP-2017-001}
placed downstream of the magnet.
The tracking system provides a measurement of the momentum, \ptot, of charged particles with
a relative uncertainty that varies from 0.5\% at low momentum to 1.0\% at 200\gev.
The minimum distance of a track to a primary $pp$ collision vertex~(PV), the impact parameter~(IP), 
is measured with a resolution of $(15+29/\pt)\mum$,
where \pt is the component of the momentum transverse to the beam, in\,\gev.
Different types of charged hadrons are distinguished using information
from two ring-imaging Cherenkov detectors~\cite{LHCb-DP-2012-003}. 
Photons, electrons, and hadrons are identified by a calorimeter system consisting of
scintillating-pad and preshower detectors, an electromagnetic
and a hadronic calorimeter. Muons are identified by a
system composed of alternating layers of iron and multiwire
proportional chambers~\cite{LHCb-DP-2012-002}.
The online event selection is performed by a trigger system ~\cite{LHCb-DP-2012-004}, 
which consists of a hardware stage, based on information from the calorimeter and muon
systems, followed by a software stage, which applies a full event
reconstruction.

At the hardware trigger stage, events are required to have a muon with high \pt or a hadron, photon or electron with high transverse energy in the calorimeters. For
hadrons, the transverse energy threshold is 3.5\gev.
  The software trigger requires a two-, three- or four-track
  secondary vertex with a significant displacement from any primary
  $pp$ interaction vertex. At least one charged particle
  must have a transverse momentum $\pt > 1.6\gev$ and be
  inconsistent with originating from any PV.
  A multivariate algorithm~\cite{BBDT,LHCb-PROC-2015-018} is used for
  the identification of secondary vertices consistent with the decay
  of a \bquark hadron.

Simulated events are used to develop the selection requirements, validate fit models,
and to evaluate the efficiencies needed for the inclusive branching fraction measurement and the amplitude analysis.
In the simulation, $pp$ collisions are generated using
  \pythia~\cite{Sjostrand:2007gs} 
  with a specific \lhcb configuration~\cite{LHCb-PROC-2010-056}.
  Decays of unstable particles
  are described by \evtgen~\cite{Lange:2001uf}, in which final-state
  radiation is generated using \photos~\cite{davidson2015photos}.
  The interaction of the generated particles with the detector, and its response,
  are implemented using the \geant
  toolkit~\cite{Allison:2006ve, *Agostinelli:2002hh} as described in
  Ref.~\cite{LHCb-PROC-2011-006}. 
  The underlying $pp$ interaction is reused multiple times, each with an independently generated signal decay~\cite{LHCb-DP-2018-004}.

\section{Selection of $\boldsymbol{\Bd \to \proton\antiproton \Kp\pim}$ candidates}
\label{Selection}
Candidate $\Bz$ mesons are reconstructed in the $\Bd \to \proton\antiproton \Kp\pim$ decay over the full $m_{\proton\antiproton}$ range. The signal channel $\Bd \to \etac \Kp\pim$ and the normalisation channel $\Bd \to \jpsi \Kp\pim$ are obtained by requiring $m_{\proton\antiproton}$ to lie within their respective charmonium-resonance regions, 2908--3058~\mev for the \etac meson and 3072--3122~\mev for the \jpsi meson.
An initial offline selection is applied to reconstructed particles. Final-state particles are required to have $\pt > 300\mev$, be inconsistent with originating from any PV, lie within the acceptance of the RICH detectors ($2.0<\eta<4.9$) and be compatible with the appropriate proton, kaon or pion mass hypotheses. Furthermore, proton candidates are required to have $10<\ptot<150\gev$, while pion and kaon candidates must have $3<\ptot<150 \gev$.
Candidate \Bz mesons are required to have a small \chisqip with respect to a PV, where \chisqip is defined as the change in the vertex-fit \chisq of a given PV when reconstructed with and without including the candidate under consideration. The PV providing the smallest \chisqip value is associated to the \Bz candidate, which is also required to be consistent with originating from this PV by applying a criterion on the angle between its momentum vector and the displacement vector from the PV to its decay vertex. 
To enhance the resolution of kinematic quantities, such as  the diproton mass distribution $m_{\proton\antiproton}$ and the two-particle mass combinations that define the DP, a kinematic fit is performed~\cite{Hulsbergen:2005pu}. In this fit, the \Bz candidate is constrained to originate from its associated PV, and its reconstructed mass is constrained to the known \Bz mass~\cite{PDG2024}.

A boosted decision tree~(BDT) algorithm~\cite{AdaBoost,Breiman} is used to suppress combinatorial background, corresponding to \Bz candidates formed by uncorrelated particles. Simulated samples of $\Bd  \rightarrow \proton \antiproton  \Kp  \pim  $ decays, generated uniformly within the available phase space of the decay, are taken as a signal proxy. The data in the mass $m_{\proton\antiproton \Kp \pim }$ region 5450--5550\mev is used as a background proxy when training the BDT, to avoid regions containing partially reconstructed $b$-hadron decays. Two BDT classifiers are trained separately for Run~1 and Run~2, in order to take into account differences between the two data-taking periods.

The input variables to the BDT classifiers are: the vertex-fit \chisq, \chisqip, a discriminating variable evaluating the pointing consistency of the reconstructed \Bd
 momentum to the \Bd decay vertex, the
flight-distance significance of the reconstructed \Bz candidates; the maximum distance of closest approach between final-state particles, and the maximum and minimum \ptot and \pt of the proton and antiproton. The Run~2 BDT additionally uses particle-identification variables of the final-state particles. These variables are adjusted to match the calibration data distributions and are not included in the Run~1 BDT due to the lack of suitable calibration samples~\cite{LHCb-DP-2018-001, LHCb-PUB-2016-021}.
The selection criteria on the BDT output are chosen in order to maximise the figure of merit defined as $S/\sqrt{S+B}$, where $S$ is the observed $\Bd  \to \proton\antiproton  \Kp  \pim $ signal yield before any BDT selection multiplied by the efficiency of
the BDT requirement evaluated using simulation, while $B$ is the combinatorial background yield, both obtained in the $m_{\proton\antiproton \Kp \pim }$ region 5200--5360\mev.
Finally, the $\Dzb \to \Kp \pim $, $\Lcbar \to \antiproton \Kp \pim $, and $\phiz \to \Kp  K^-$ decays are vetoed by excluding the mass ranges 1835--1880\mev, 2250--2305\mev, and 1010--1028\mev in the $m_{\Kp \pim }$, $m_{\antiproton \Kp \pim }$, and $m_{\Kp K^-}$ distributions, respectively, where the $m_{\Kp K^-}$ mass has been reconstructed assigning the $K^-$ meson mass to the $\pim $ candidate.

\subsection{Signal yield of $\boldsymbol{\Bd \to \etac \Kp \pim}$ decay}
\label{SignalDP}
Prior to the DP fit, a two-dimensional extended maximum-likelihood fit to the $m_{\proton\antiproton K^+\pi^-}$ and $m_{\proton\antiproton}$ mass distributions is performed in order to isolate the $\Bd  \to \etac (\to \proton\antiproton)\Kp \pim $ contribution from the nonresonant~(NR) component, $\Bd  \to \proton\antiproton \Kp \pim $. The ranges of the fits are 5220--5340\mev and 2908--3058\mev for the $\proton\antiproton \Kp \pim $ and $\proton\antiproton$ mass distributions, respectively. 

\begin{figure}[tb]
\centering
\includegraphics[width=0.49\textwidth]{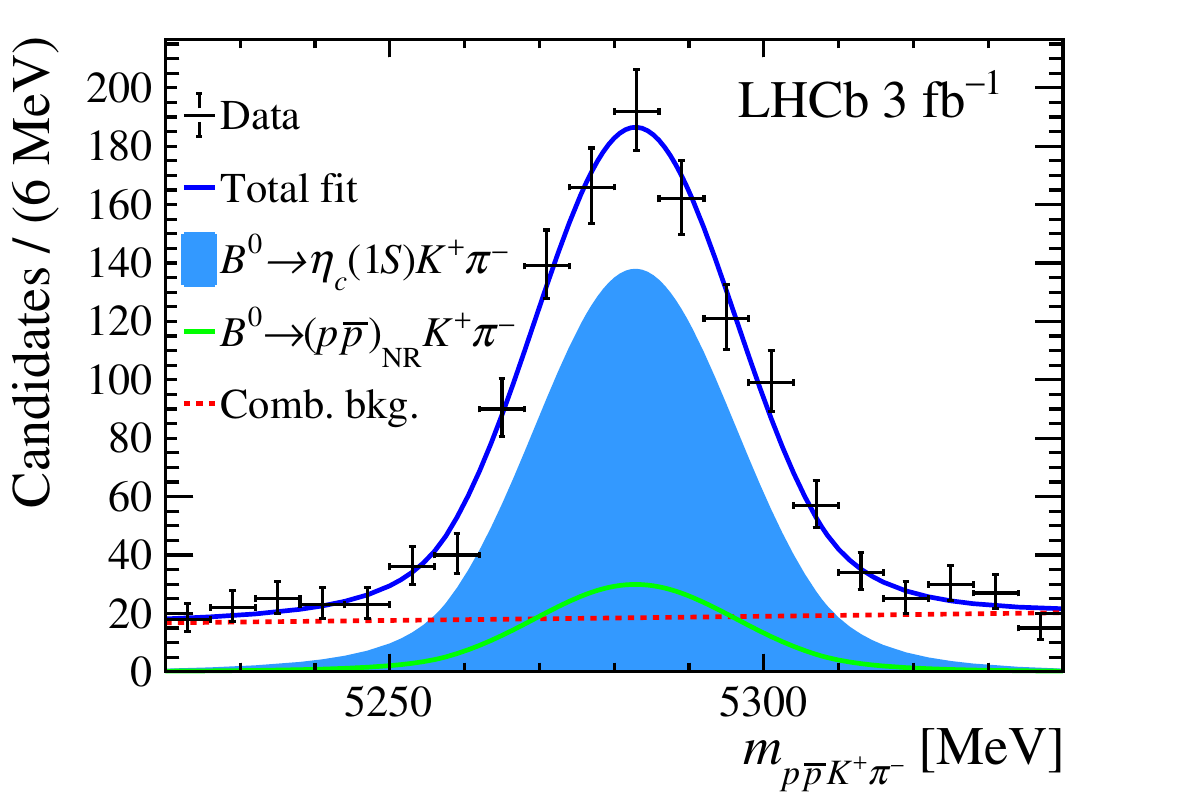}
\includegraphics[width=0.49\textwidth]{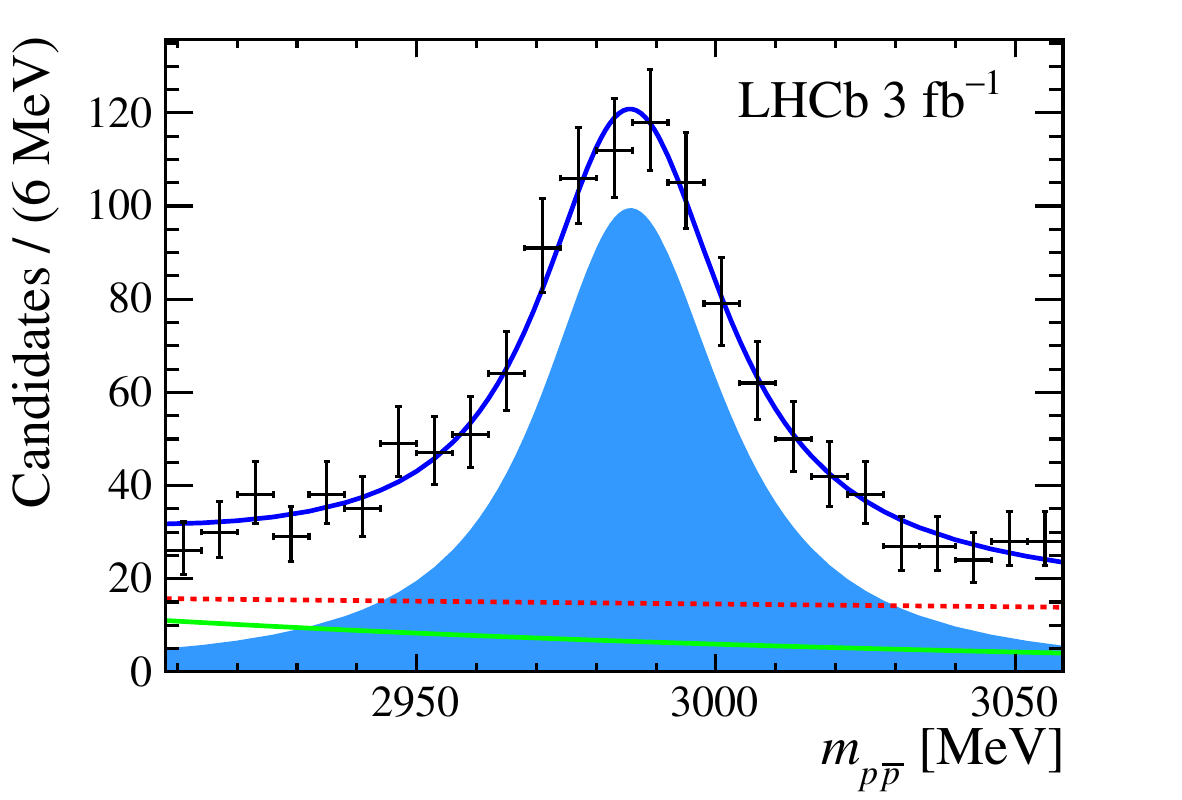}
\includegraphics[width=0.49\textwidth]{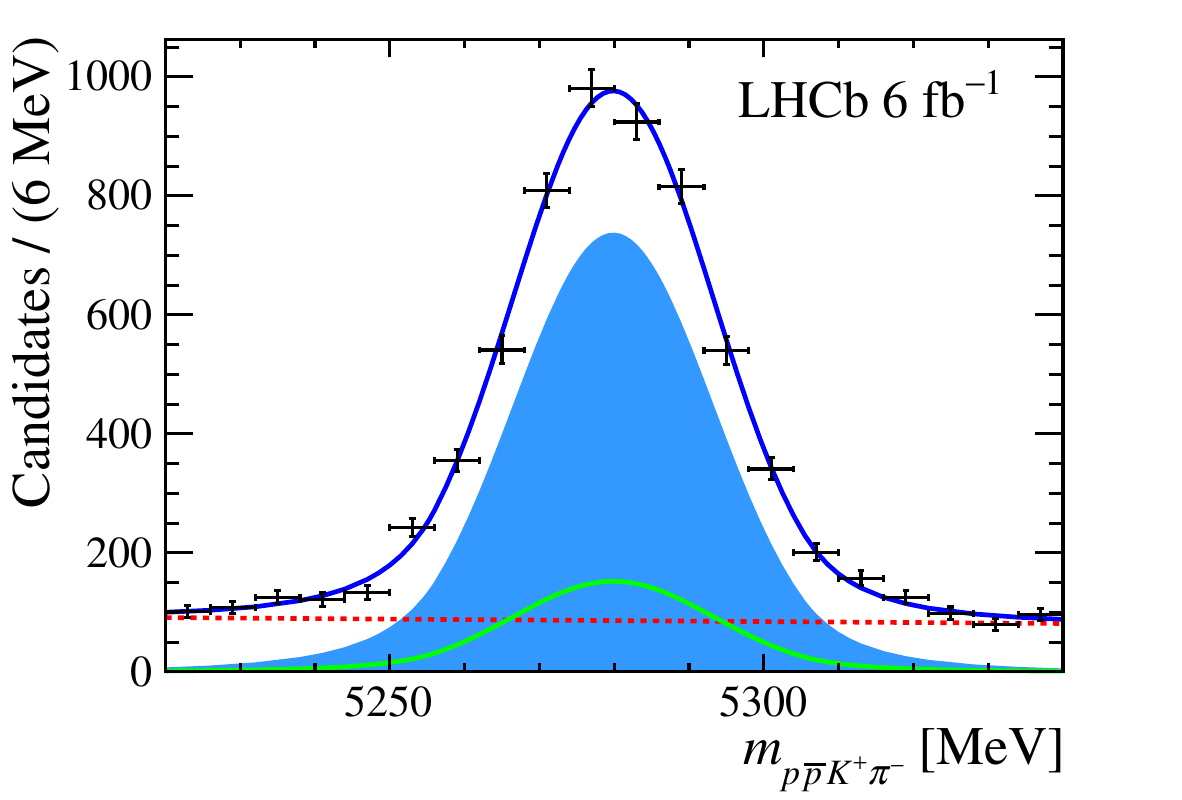}
\includegraphics[width=0.49\textwidth]{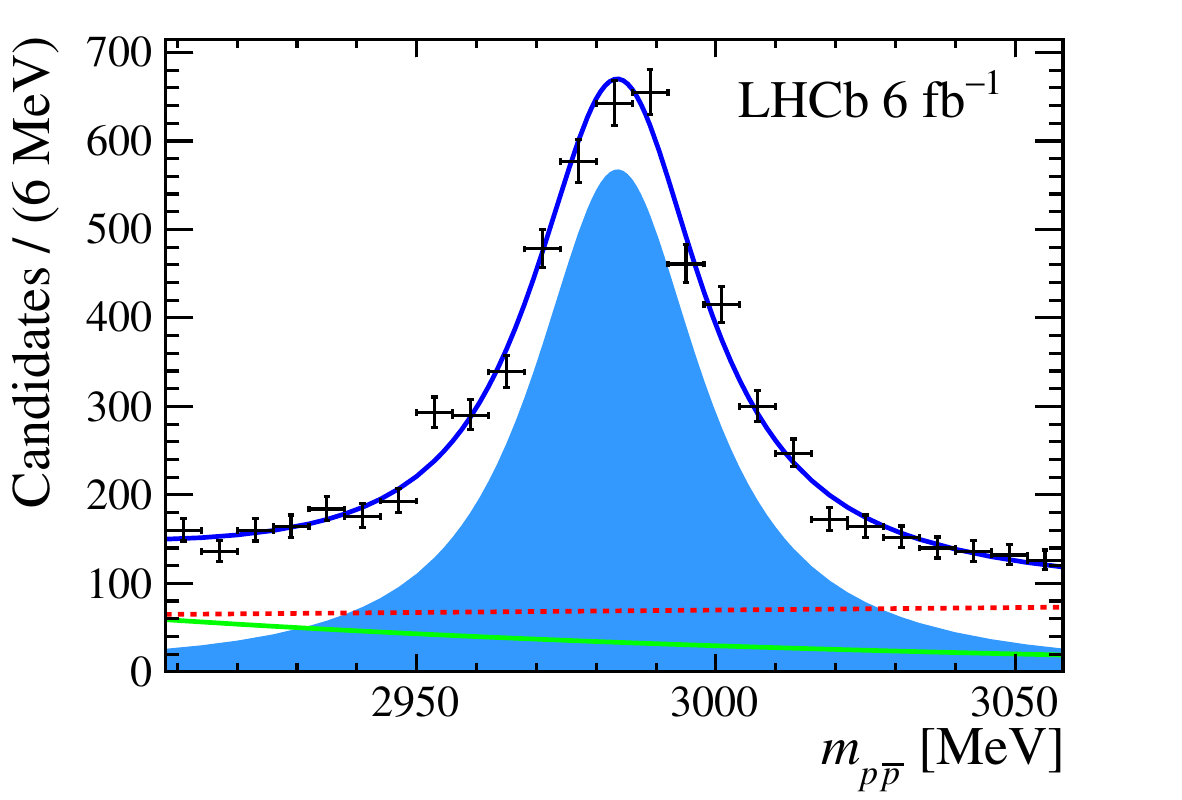}
\caption{Results of the two-dimensional mass fit to the (left)~$m_{\proton\antiproton \Kp \pim }$ and (right)~$m_{\proton\antiproton}$ distributions for the (top)~Run~1 and (bottom) Run~2 datasets in the region of the \etac resonance.
}
\label{2DFit}
\end{figure}

The \roofit package~\cite{roofit} is used to perform the fit separately for the Run~1 and Run~2 data samples.
The $m_{\proton\antiproton \Kp \pim }$ distributions of $\Bd \to{\etac \Kp \pim }$ signal and $\Bd \to{\proton\antiproton \Kp \pim }$ NR decays are both described by a Hypatia probability density functions (PDF)~\cite{Santos:2013gra} using the same parametrisation. 
The $m_{\proton\antiproton}$ mass distribution of the $\Bd \to \etac \Kp \pim $ signal decays is described by a relativistic Breit--Wigner (RBW) convolved with a Crystal Ball function~\cite{Skwarnicki:1986xj}, while that of the NR contribution is described by an exponential function. 
The $m_{\proton\antiproton \Kp \pim }$ and $m_{\proton\antiproton}$ distributions of the combinatorial background are modelled using exponential functions.
A possible component where $\etac $ mesons are combined with random kaons and pions from the PV is investigated but found to be negligible. 
The meson peak positions,
widths, slopes of the exponential functions, and yields, are free to vary in the mass fits while the shape parameters of the signal component are fixed from simulation. 
The yields of all fit components are reported in Table~\ref{Fit2Dres}. Figure~\ref{2DFit} shows the projections of the mass fits for the Run~1 and Run~2 samples.

A fit to the $m_{\proton\antiproton}$ distribution is also performed incorporating a possible interference term between the NR and the $\etac$ components. This term is found to be consistent with zero and is therefore neglected in the subsequent analysis.
In order to test if local interferences are present, the $m_{\proton \antiproton}$ distribution is studied in different bins of helicity angles of the $\Kp \pim$ system. No evidences of local interferences are observed.
{ 
\begin{table}[tb]
\begin{center}
\caption{Yields of the signal and background components from the two-dimensional mass fit to the $m_{\proton\antiproton \Kp \pim }$ and $m_{\proton\antiproton}$ distributions for the Run~1 and Run~2 samples in the \proton\antiproton mass region of the \etac resonance.}
\begin{tabular}{l r r} 

 \hline
 & \multicolumn{2}{c}{Yields}\\
 \hline
 Component & Run~1\phantom{0} & Run~2\phantom{00} \\
 \hline
    $\Bd \to{\etac \Kp \pim }$ & 772 $\pm$ 70 & 4266 $\pm$ 139\\
    $\Bd \to{\proton\antiproton \Kp \pim }$ & 168 $\pm$ 69 & 823 $\pm$ 139\\
    Combinatorial background & 361 $\pm$ 34& 1665 $\pm$ \phantom{0}68\\
 \hline

\end{tabular}

\label{Fit2Dres}
\end{center}
\end{table}
}

\subsection{Signal yield of the $\boldsymbol{\Bd\to\jpsi\Kp\pim}$ decay}
\begin{figure}[tb]
\centering
\includegraphics[width=0.49\textwidth]{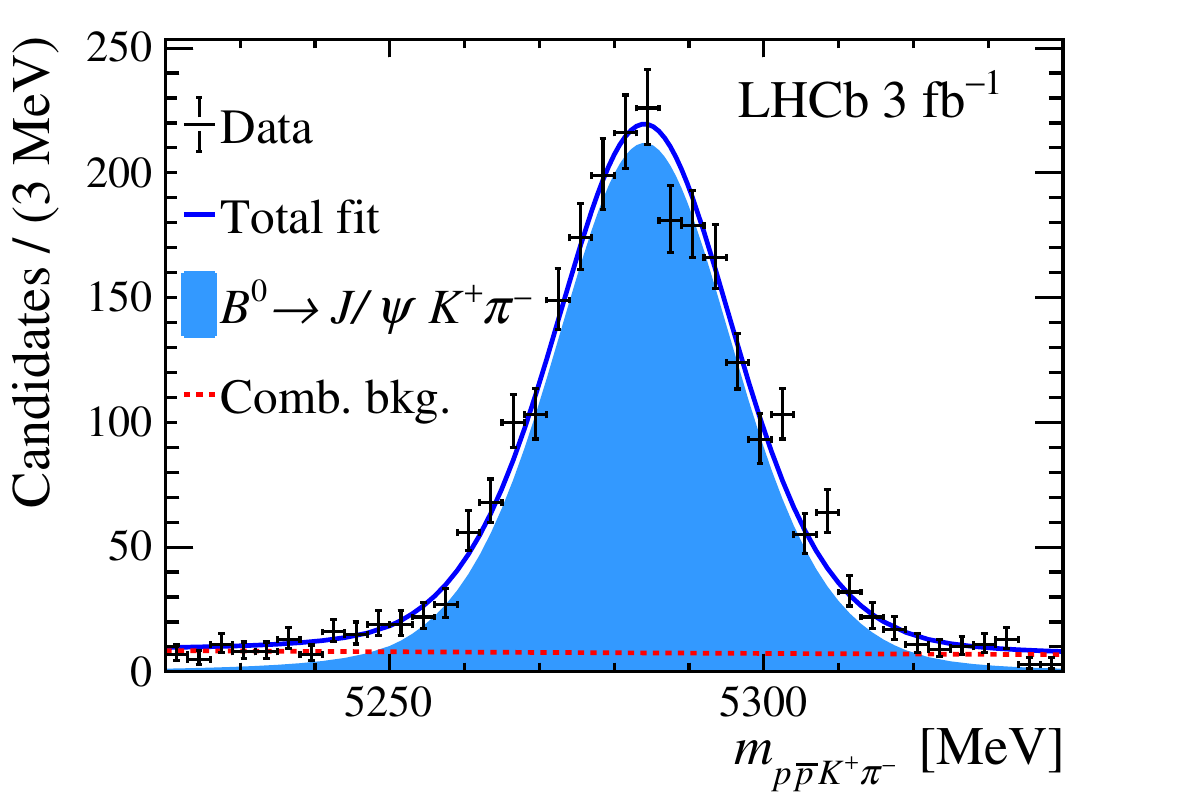}
\includegraphics[width=0.49\textwidth]{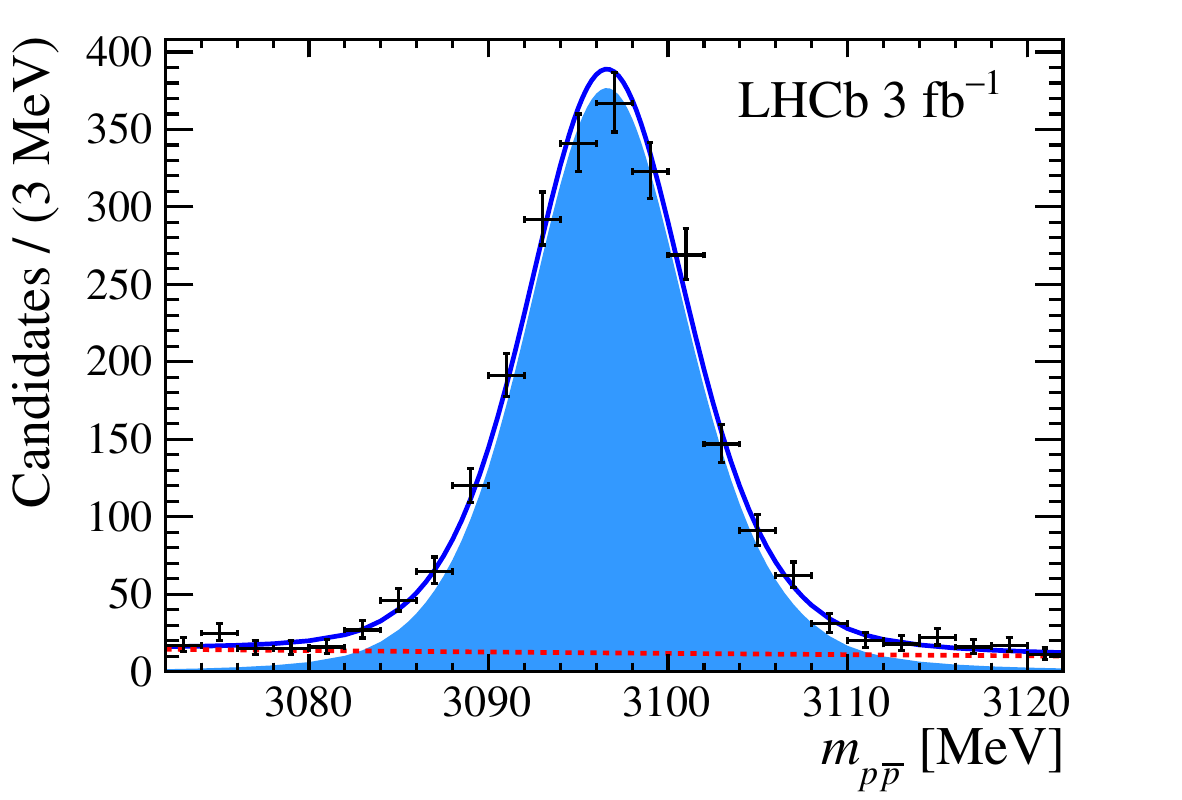}
\includegraphics[width=0.49\textwidth]{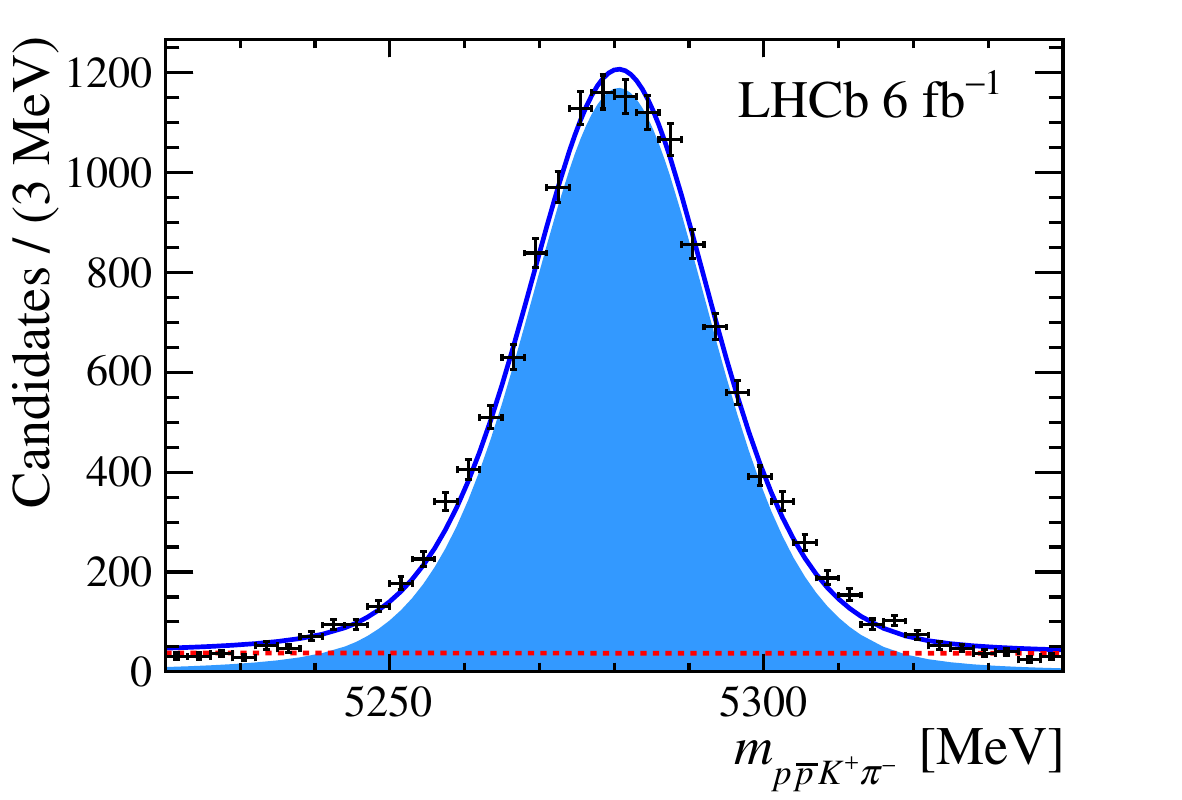}
\includegraphics[width=0.49\textwidth]{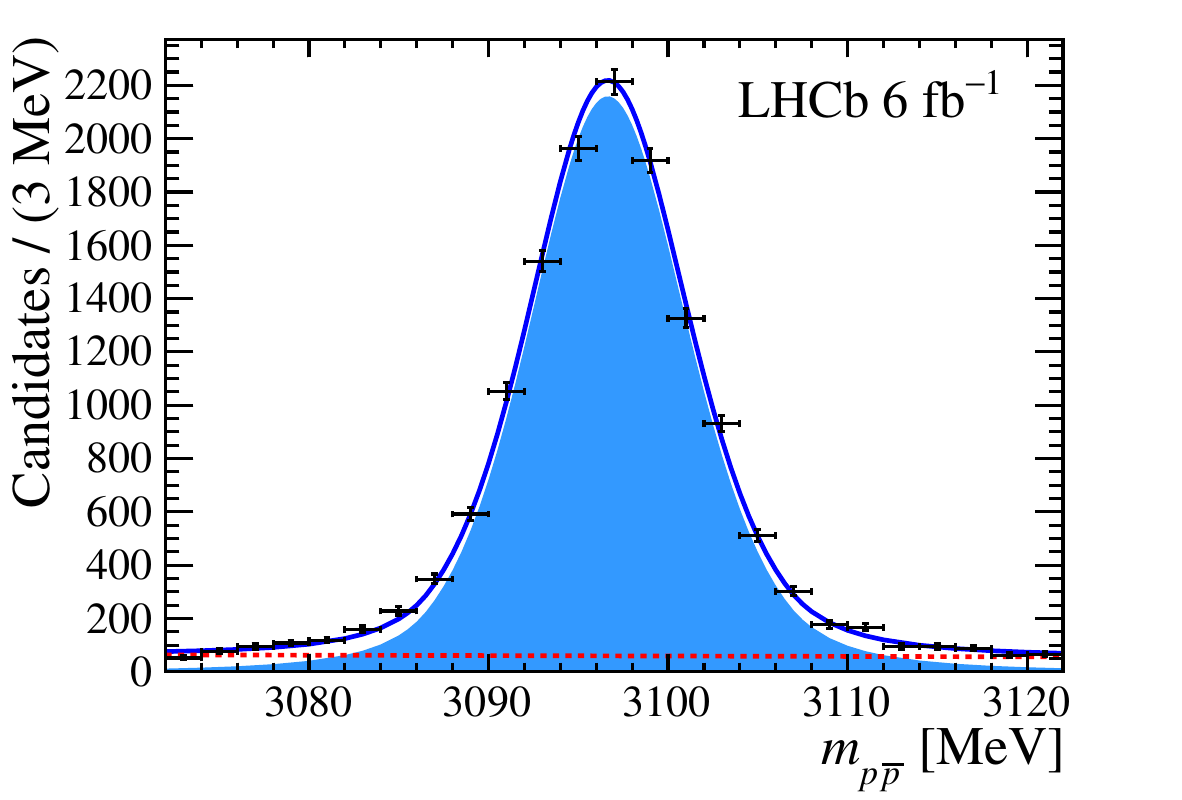}
\caption{Results of the two-dimensional fit to the (left)~$m_{\proton\antiproton \Kp \pim }$ and (right)~$m_{\proton\antiproton}$ distributions for the (top)~Run~1 and (bottom) Run~2 datasets in the region of the \jpsi resonance.}

\label{2DFitjpsi}
\end{figure}

A similar two-dimensional fit procedure is also used to extract the yield of $\Bd \to{ \jpsi \Kp \pim }$ decays.
 A Hypatia PDF is used to parametrise the $ \jpsi \to \proton\antiproton$ decay in the range \mbox{3072--3122~\mev}, with the shape parameters fixed from simulation. Nonresonant contributions from $\mbox{\Bd \to{\proton\antiproton \Kp \pim }}$ decays are not included in the fit that is used to measure $N_{\jpsi}$, as they are found to be insignificant. The yields of the fit components are reported in Table~\ref{Jpsiyield}, while the results of the mass fits in the $m_{\proton\antiproton}$ mass region of the \jpsi resonance is shown in Fig.~\ref{2DFitjpsi}.
{ 
\begin{table}[tb]
\begin{center}
\caption{Yields of the normalisation and background components from the two-dimensional joint mass fit to the $m_{\proton\antiproton \Kp \pim }$ and $m_{\proton\antiproton}$ distributions for the Run~1 and Run~2 samples in the \proton\antiproton mass region of the \jpsi resonance.}
\begin{tabular}{  l c c  }
\hline
 & \multicolumn{2}{c}{Yields}\\
 \hline
 Component & Run~1 & Run~2 \\
 \hline
    $\Bd \to{ \jpsi \Kp \pim }$ & 2258 $\pm$ 50 & 12787 $\pm$ 119\\
    Combinatorial background & \phantom{0}306 $\pm$ 24& \phantom{0}1496 $\pm$ \phantom{0}54\\
 \hline

\end{tabular}

\label{Jpsiyield}
\end{center}
\end{table}
}

\section{Dalitz plot fit}
\begin{figure}[tb]
\centering
\includegraphics[width=0.96\textwidth]{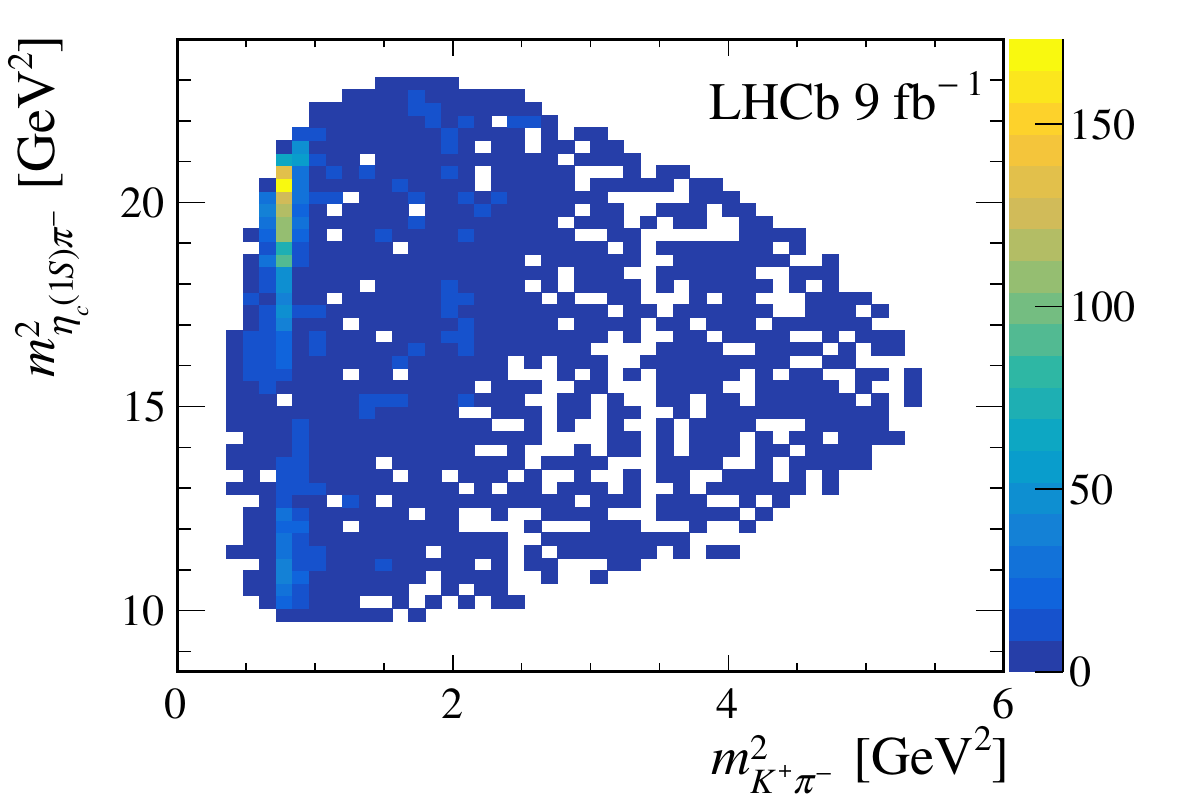}
\caption{Background-subtracted DP distribution of the $\Bd \to \eta_c \Kp \pim$ decay.}
\label{DP}
\end{figure}

Using the \sPlot technique~\cite{Pivk:2004ty} applied to the results of the two-dimensional mass fit in the region of the \etac resonance, the background-subtracted DP is determined and shown in Fig.~\ref{DP}.
The phase space for a three-body decay involving only pseudoscalar particles can be represented by a DP, where two of the three possible two-body mass-squared combinations, in this case $m^2_{\Kp \pim }$ and $m^2_{\etac \pim }$, define the DP axes. Due to the sizeable natural width of the $\etac $ meson, the $p$ and $\antiproton$ momenta and energies are used instead of fixing the $\etac $ mass to its known value~\cite{PDG2024} to compute kinematic quantities such as $m^2_{\etac  \Kp }$, $m^2_{\etac \pim }$, and the helicity angles.

The formalism used in this analysis follows the approach described in Ref.~\cite{LHCb-PAPER-2018-034} where an unbinned maximum-likelihood fit on the DP is performed employing the $\textsc{Laura++}$ package~\cite{Back:2017zqt}. The Run~1 and Run~2 subsamples are fitted simultaneously within the JFIT framework~\cite{Ben-Haim:2014afa}, with shared free parameters in the DP fit; however, the signal and background yields, as well as the efficiency-variation maps across the phase space, are treated separately for the two datasets.

In the DP fit, the signal corresponds to $\Bd  \to \etac  \Kp  \pim $ decays, while the background includes both combinatorial contributions and NR contributions. The likelihood function is given by
\begin{equation}
\mathcal{L}=  e^{-\sum_k N_k}  \prod_i\left[  \sum_k   N_{k}   \mathcal{P}_{k}  (  m_{\Kp\pim }^ {2}  ,  m_{\etac  \pim}^ {2})_{i}\right],
\end{equation}
where the index $ i $ runs over the number of candidates, while $ k $ indexes the signal and background components, with $ N_k $ representing their respective yields. The PDFs for the signal and background components are denoted by $\mathcal{P}_k$, for which the signal PDF is defined as the normalised squared amplitudes scaled by an efficiency model in the DP.
Combinatorial and NR background shapes are extracted from the mass fits using the \sPlot method. These histograms are also interpolated with a cubic spline before the DP fit. 

To mitigate issues arising from imperfect efficiency parametrisation at the DP boundaries, a veto is applied to both the lower and upper edges of the DP, effectively excluding regions where any of the masses, $m_{\etac \pim}$, $m_{\etac \Kp}$, or $m_{\Kp \pim}$, are within $70$\mev of their kinematic limits.
This veto is also used in both the determination of signal and background yields and the PDF modelling the background.  

The mass resolution of the $\Kp \pim  $ system is approximately 5\mev, which is significantly smaller than the natural width of the narrowest resonance in the Dalitz plot, the $\Kstar(892)^0 $ meson, whose width is about 50\mev. Therefore, resolution effects are negligible and are not considered further. To ensure convergence to the global minimum, the DP fits are repeated multiple times with randomised initial parameter values.

\subsection{Signal efficiency evaluation}
\label{efficiency}

In order to perform the DP fit of the $\Bd  \rightarrow \etac  \Kp  \pim $ decay channel, the efficiency variations across the phase space of the decay must be taken into account. 
Relative variations in efficiency arise from both experimental effects and selection requirements. Efficiencies are calculated on a two-dimensional map parametrised by the variables $m^\prime$ and $\theta^\prime$ which are defined as
\begin{equation}
m^{\prime} \equiv \frac{1}{\pi} \arccos{ \left (2 \frac{m_{\Kp \pim }-m^{\mathrm{min}}_{\Kp \pim }}{m^{\mathrm{max}}_{\Kp \pim }-m^{\mathrm{min}}_{\Kp \pim }}-1 \right )},
\end{equation}
\begin{equation}
\theta^{\prime} \equiv \frac{1}{\pi} \theta_{\Kp \pim },
\end{equation}
where the helicity angle $\theta_{\Kp \pim }$ is defined as the angle between the $\Kp $ and the $\etac $ particles in the
$ \Kp \pim $ rest frame. This parametrisation is used to obtain a square Dalitz plot which mitigates potential issues related to the otherwise curved boundaries in the $m_{\Kp \pim }^2$ and $m_{\etac  \pim }^2$ space and the finite width of the $\etac $ resonance. The phase-space boundaries of the $\Kp \pim $ mass are represented by $m^{\mathrm{min}}_{\Kp \pim }$ and $m^{\mathrm{max}}_{\Kp \pim }$.

The efficiency distributions are computed using simulation as two-dimensional histograms in the square DP space, using the four-momenta of the final-state particles. 
These momenta are obtained from a fit with the selected \Bz candidates, where the \Bz mass is constrained to its known value~\cite{PDG2024} and its direction required to point towards the associated primary vertex.
The efficiency distributions are calculated by applying the full event selection, where potential discrepancies arising from trigger, tracking, and \Bz-kinematic sources are found to be negligible. The effect of the vetoes in the phase space is separately accounted for by the $\textsc{Laura++}$ package, setting the signal efficiency to zero both within the excluded regions and beyond the phase-space border restrictions. The histograms modelling the efficiency undergo a smoothing procedure through cubic-spline interpolation in order to reduce the statistical fluctuations due to the limited size of the simulated samples and are studied separately for the Run~1 and Run~2 samples.

\subsection{\boldmath Amplitude model with only \Kstarz resonances}

{ 
\begin{table}[tb]
\begin{center}
\caption{Resonances included in the baseline model, where parameter values, uncertainties and spin-parity $J^P$ are taken from Ref.~\cite{PDG2024}. The lineshapes used are also reported.
}
\begin{tabular}{  l r r c c }

 \hline
 $K^{*0}$ resonance & Mass [\mev] & Width [\mev] & $J^P$ & Lineshape\\
 \hline
 $K^*(892)^0$& 895.55 $ \pm $ \phantom{0}0.20& 47.3 $ \pm $ \phantom{00}0.5 &$1^-$ &RBW \\ 
 $K^*(1410)^0$& 1414\phantom{.00} $ \pm $ 15\phantom{.00}  & 232\phantom{.0}  $ \pm $\phantom{0} 21\phantom{.0} & $1^-$ & RBW\\
 $K^*_0(1430)^0$& 1425\phantom{.00} $ \pm $ 50\phantom{.00} & 270\phantom{.0} $ \pm $ \phantom{0}80\phantom{.0}&$0^+$ & LASS\\
 $K^*_2(1430)^0$& 1432.4\phantom{0} $ \pm $ \phantom{0}1.3\phantom{0} & 109\phantom{.0} $ \pm $ \phantom{00}5\phantom{.0} & $2^+$& RBW\\
 $K^*(1680)^0$& 1718\phantom{.00} $ \pm $ 18\phantom{.00} & 320\phantom{.0} $ \pm $ 110\phantom{.0}&$1^-$ & RBW\\
 $K^*_0(1950)^0$& 1957\phantom{.00} $ \pm $ 14\phantom{.00}& 170\phantom{.0} $ \pm $ \phantom{0}50\phantom{.0}& $0^+$& RBW\\
 \hline

\end{tabular}

\label{resonanceused}
\end{center}
\end{table}
}

\begin{table}[!t]
\begin{center}
\caption{Parameters of the LASS function resulting from the best fit using the baseline model, where the uncertainties are statistical only.}
\begin{tabular}{  l  c  }
\hline
 Parameter & Value\\
 \hline
 $r$ & \phantom{000}$4.0\pm\phantom{0}1.0 \gev^{-1}$\\
 $a$ & \phantom{000}$4.3\pm\phantom{0}0.8 \gev^{-1}$\\
 $m_{K^{*}_{0}(1430)}$ &  $1434\phantom{.0}\pm22\phantom{.0}$\mev\phantom{0}\\
$\Gamma_{K^{*}_{0}(1430)}$ & \phantom{0}$409\phantom{.0}\pm44\phantom{.0} $\mev\phantom{0}\\
 \hline
\end{tabular}
\label{LASSdef}
\end{center}
\end{table}

\begin{figure}[tb]
\centering
\includegraphics[width=0.49\textwidth]{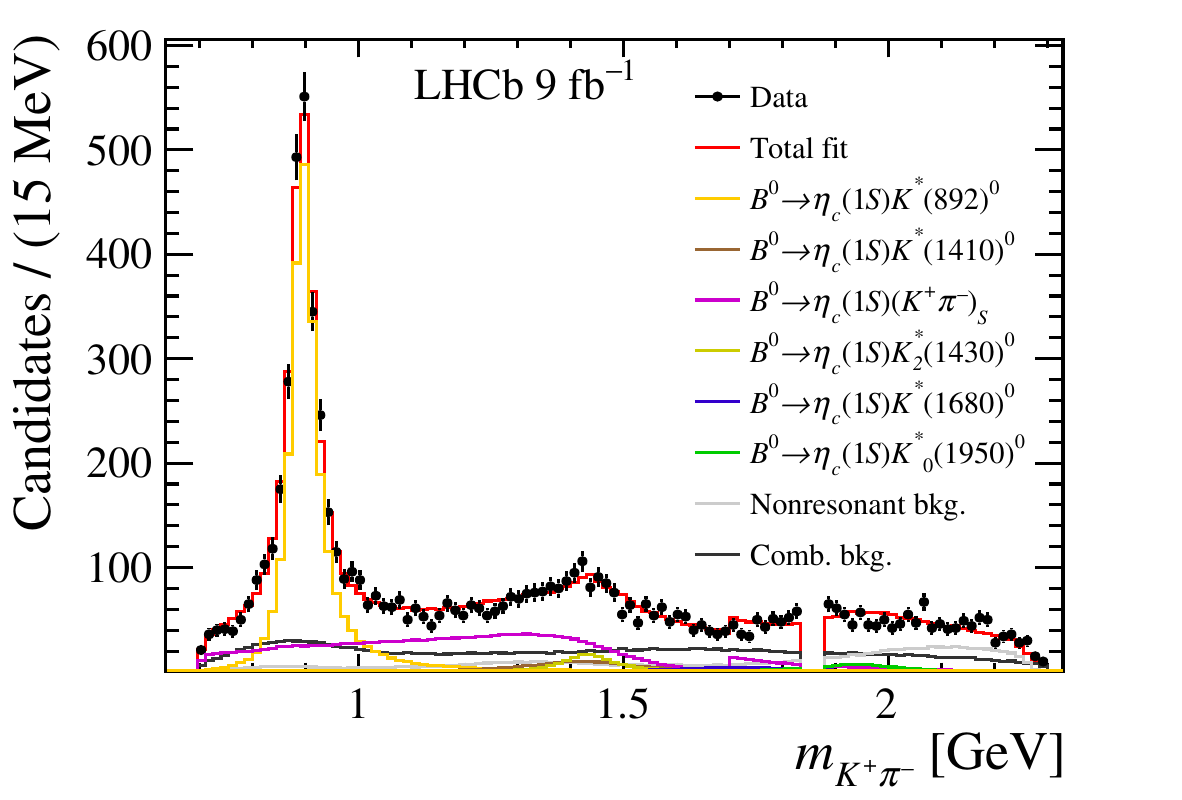}
\includegraphics[width=0.49\textwidth]{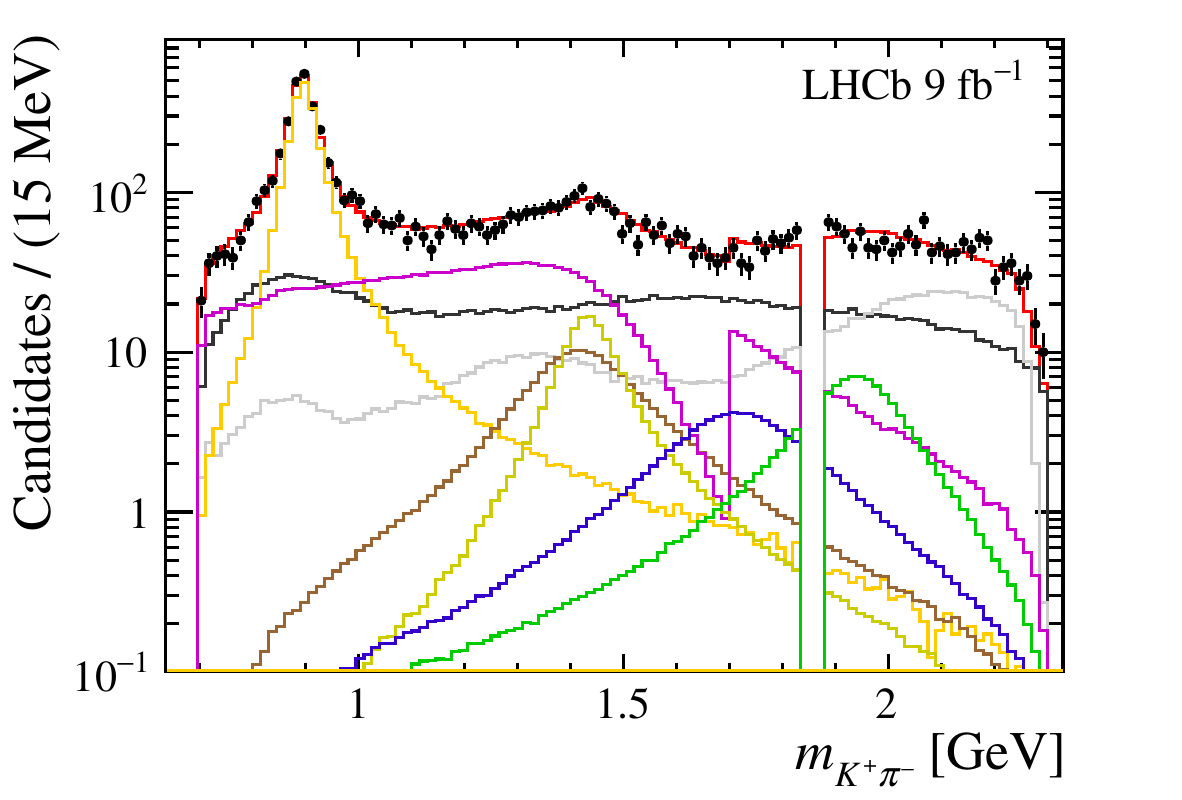}
\includegraphics[width=0.49\textwidth]{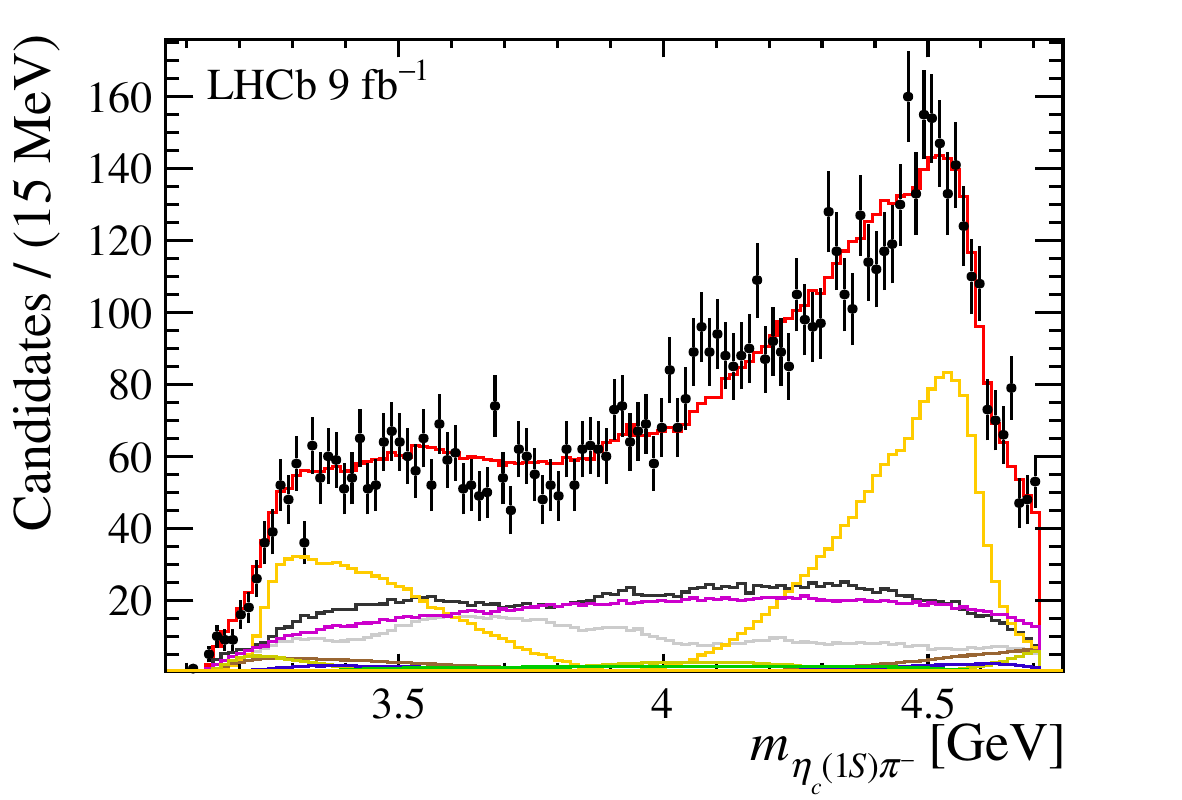}
\includegraphics[width=0.49\textwidth]{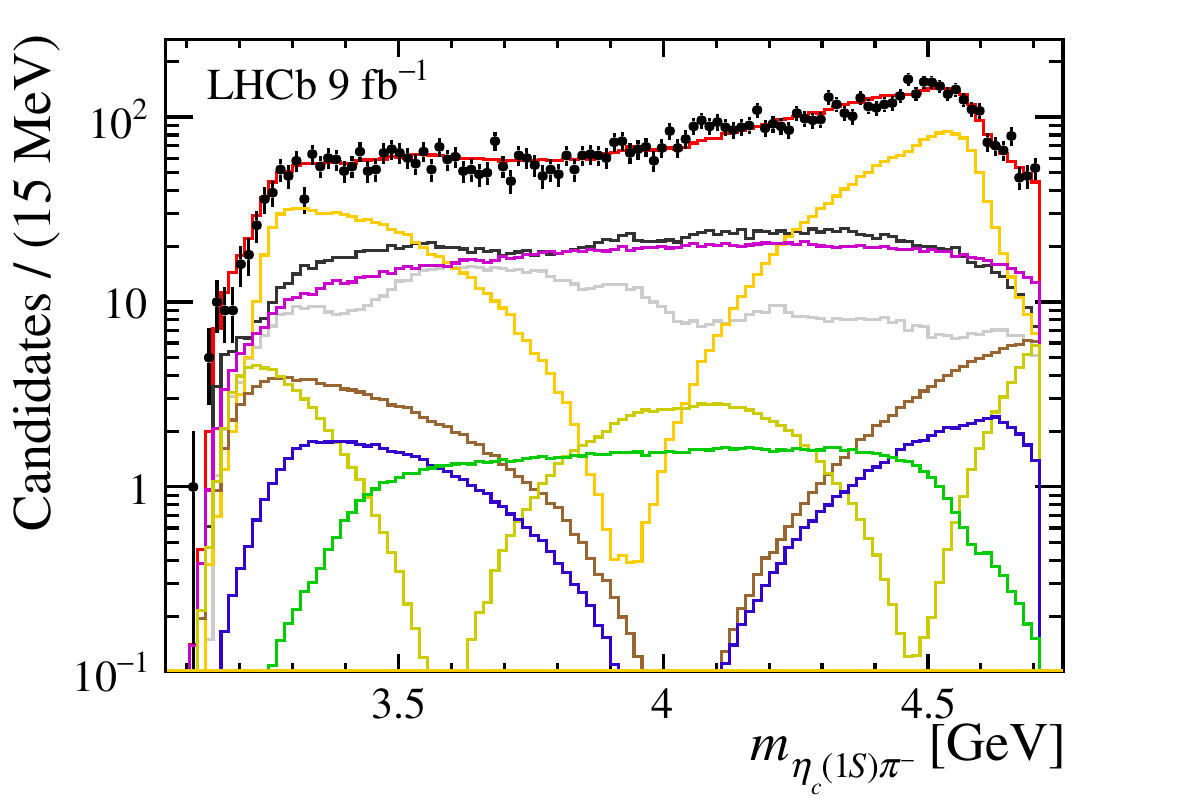}
\includegraphics[width=0.49\textwidth]{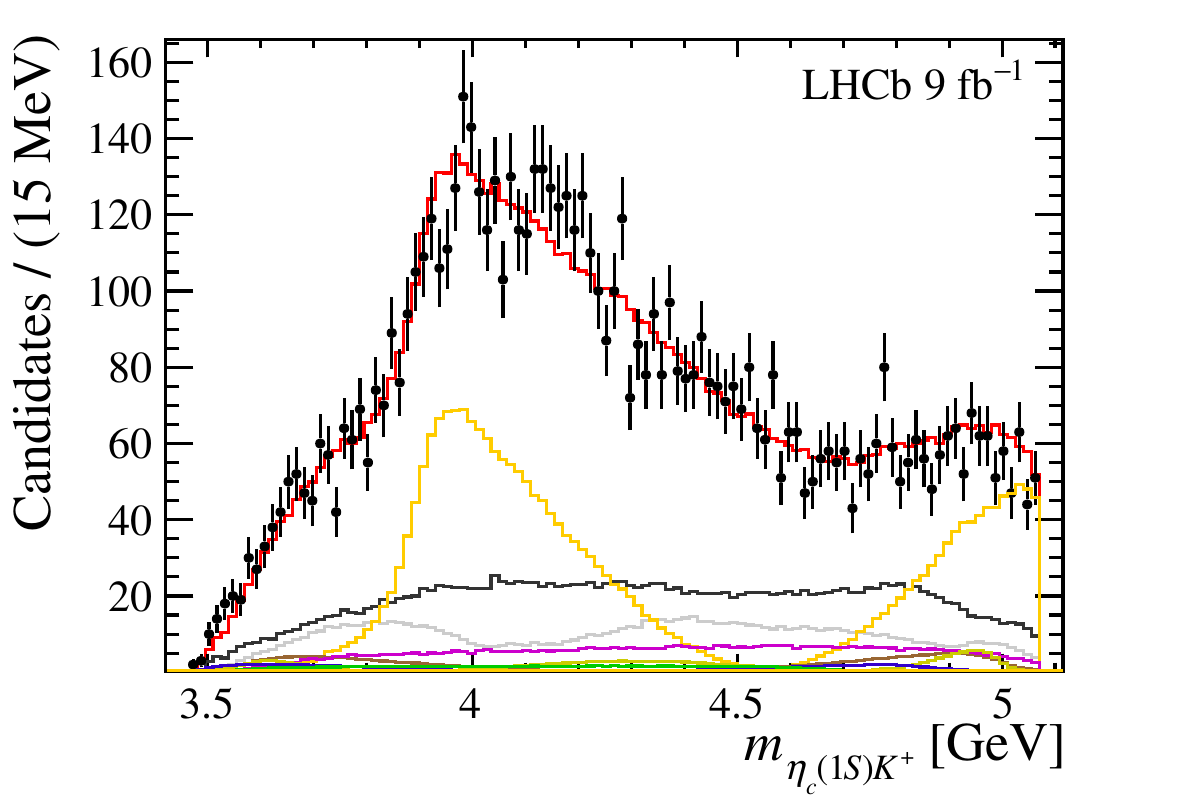}
\includegraphics[width=0.49\textwidth]{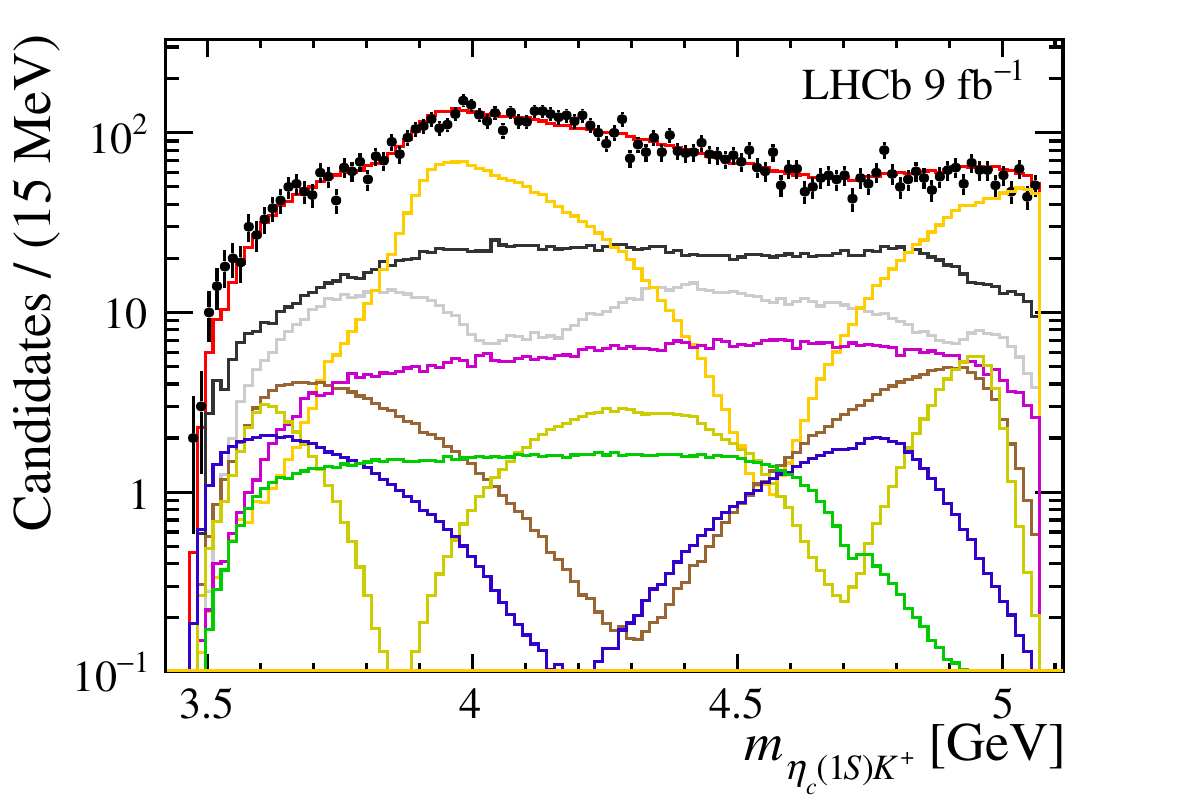}
\caption{Projections of the $\Bd\to\etac\Kp\pim$ data and DP fit using the baseline model onto the (top)~$m_{\Kp \pim }$, (middle)~$m_{\etac \pim }$ and (bottom)~$m_{\etac \Kp }$ observables for (left) linear and (right) logarithmic vertical-axis scale. The veto of $\Bd  \to \proton \antiproton \Dzb$ decays is visible in the projections onto the~$m_{\Kp \pim }$ observable.}
\label{DPfitDefault}
\end{figure}
The established $\Kstarz\to \Kp \pim $ mesons, reported in Ref.~\cite{PDG2024}, satisfying $m_{K^{*}}\lesssim m_{\Bd }-m_{\etac }$, \ie with masses within or slightly above the phase-space
boundary in $\Bd  \to \etac  \Kp  \pim $ decays, serve as a guide when building the model. Only contributions that significantly improve the data description are included. This model, referred to as the baseline model, comprises the resonances shown in Table~\ref{resonanceused}, with their masses and widths fixed to their known values~\cite{PDG2024}, except for the $K^*_0(1430)^0$ as described later. 
In the DP fit, the signal and background yields are fixed to the values obtained from the two-dimensional mass fit. The effect of the phase-space boundary cut on these yields is negligible.

The $\Kstar(892)^0$ resonance serves as a reference amplitude, with its magnitude fixed to unity and  phase set to zero. The complex coefficients~\cite{PhysRevD.11.3165,PhysRev.166.1731,PhysRev.135.B551} of the remaining contributions are determined relative to those of the $K^{*}(892)^0$ meson.
The low-mass $(\Kp\pim)_S$ S-wave is modelled with the LASS function~\cite{Aston:1987ir}. The scattering length~($a$) and the effective range~($r$) parametrising the slowly varying part~(SVP) of the LASS function are left free to vary in the fit. The resonance mass and width are constrained with Gaussian priors centred on their known values~\cite{PDG2024}. Once the global minimum of the negative log-likelihood is reached, the four parameters reported in Table~\ref{LASSdef} are fixed to their best-fit values and then the fits are repeated. This procedure negates the recalculation of resonant lineshapes in the function minimisation loop, ensuring faster convergence. The remaining resonances in the fit are modelled using RBW lineshapes, with their Blatt--Weisskopf barrier-factor radii fixed at 4$\gev^{-1}$. The addition of the $K_3^*(1780)$ and $K_4^*(2045)$ mesons are found to be not significant and therefore not included in the model. 

A good description of the $m_{\Kp \pim }$ and $m_{\etac  \Kp }$ mass distributions with only \Kstarz contributions is generally achieved. However, discrepancies between data and fit projections onto the $m_{\etac  \pim }$ and $m_{\etac  \Kp }$ distributions are visible around
$m_{\etac  \pim } \sim 4.1$\gev and $m_{\etac  \Kp } \sim 4.1$\gev, as shown in Fig.~\ref{DPfitDefault}. The values of the magnitudes, phases and fit fractions for each contribution are reported in Table~\ref{nominalfit}. A \chisq variable is computed on the squared DP to quantitatively assess the fit quality, using an adaptive two-dimensional binning scheme that ensures at least 20 events per bin, corresponding to 324 bins. For the baseline model and binning scheme, the goodness of the fit is \chisqndf = 425/309, where ndf is the number of degrees of freedom.

Two additional validation checks are performed. First, a mixed-sample unbinned goodness-of-fit test~\cite{Williams:2010vh} is used to remove binning effects; the fit to the resulting test statistic yields \chisqndf = 22.6/21, consistent with good agreement between data and model. Second, 1000 pseudoexperiments are generated and fitted with the nominal procedure. For each pseudoexperiment a \chisqndf value is computed as described above, and the p-value is defined as the fraction of pseudoexperiments with \chisqndf larger than that observed in data. Neglecting systematic uncertainties, this procedure gives a p-value of 3.1\%.

\begin{table}[tb]
\begin{center}
\caption{Magnitudes, phases and fit fractions determined from the DP fit using the baseline model,  where the uncertainties are statistical only. }
\begin{tabular}{  l c c c  }
\hline
 Amplitude & Magnitude & Phase & Fit fraction ($\%$)\\
 \hline
 $\Bd  \to \etac \Kstar(892)^0\phantom{1s}$ & 1 (fixed) & 0 (fixed) & $\phantom{0}49.3\pm1.2$\\ 
  $\Bd  \to \etac K^*(1410)^0\phantom{s}$ & $0.30 \pm 0.04$ & $-0.14\pm0.13$& $\phantom{00}4.5\pm1.4$\\ 
  $\Bd  \to \etac K_0^*(1430)^0\phantom{s}$ & $0.79 \pm 0.04$ &$+2.94\pm0.05$& $\phantom{0}30.8\pm4.7$\\ 
    $\Bd  \to \etac  (\Kp\pim)_{\rm SVP}$ & $0.56 \pm 0.03$ & $+2.81 \pm 0.07$ & $\phantom{0}15.4\pm2.4$\\ 
 $\Bd  \to \etac K_2^*(1430)^0\phantom{s}$ & $0.28 \pm 0.03$ & $-0.73 \pm 0.10$& $\phantom{00}3.7\pm0.9$\\ 
 $\Bd  \to \etac K^*(1680)^0\phantom{s}$ & $0.21 \pm 0.05$ & $+0.21 \pm 0.22$& $\phantom{00}2.1\pm0.9$\\ 
 $\Bd  \to \etac K_0^*(1950)^0\phantom{s}$ & $0.22 \pm 0.03$ & $+1.53\pm 0.29$ & $\phantom{00}2.4\pm0.5$\\   
 \hline
 Sum of fit fractions & & & $108.2\pm3.1$\\
 \hline
\end{tabular}
\label{nominalfit}
\end{center}
\end{table}

\subsection{Amplitude model with $\boldsymbol{\Kstarz}$ resonances and an $\boldsymbol{\etac \pim }$ exotic contribution}
\begin{figure}[!h]
\centering
\includegraphics[width=0.49\textwidth]{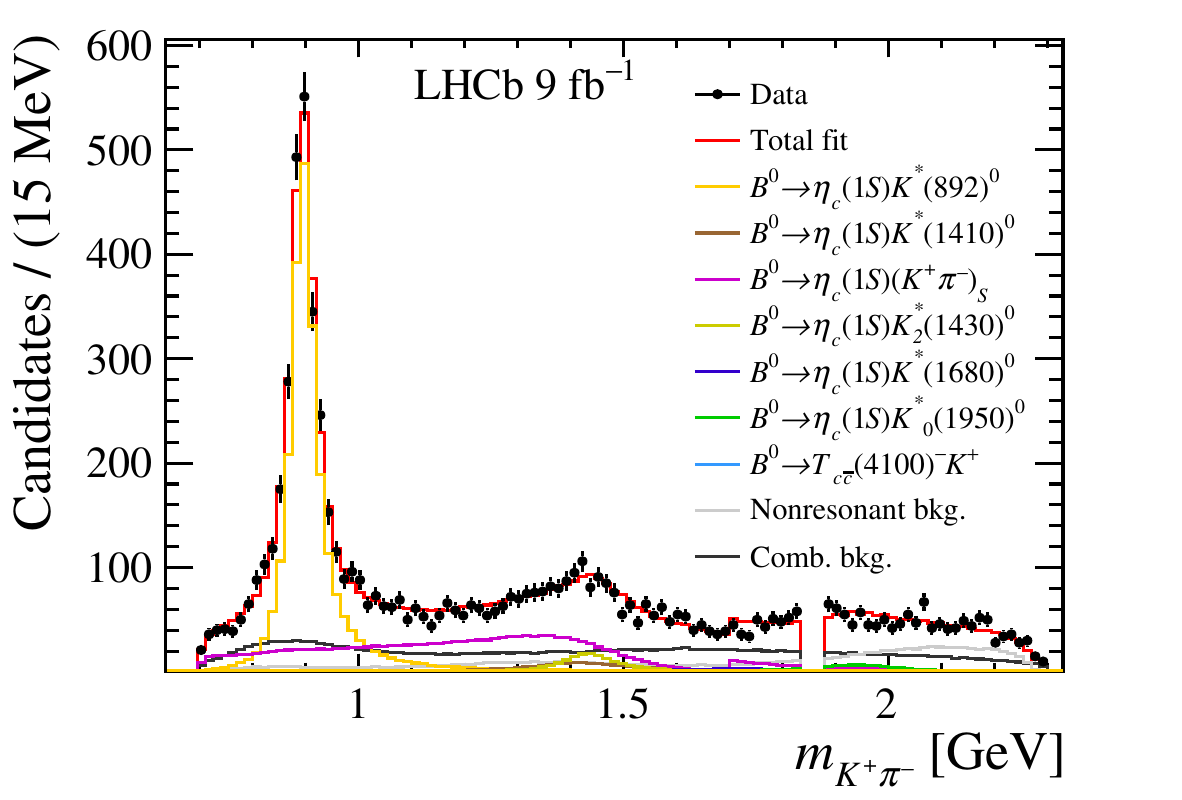}
\includegraphics[width=0.49\textwidth]{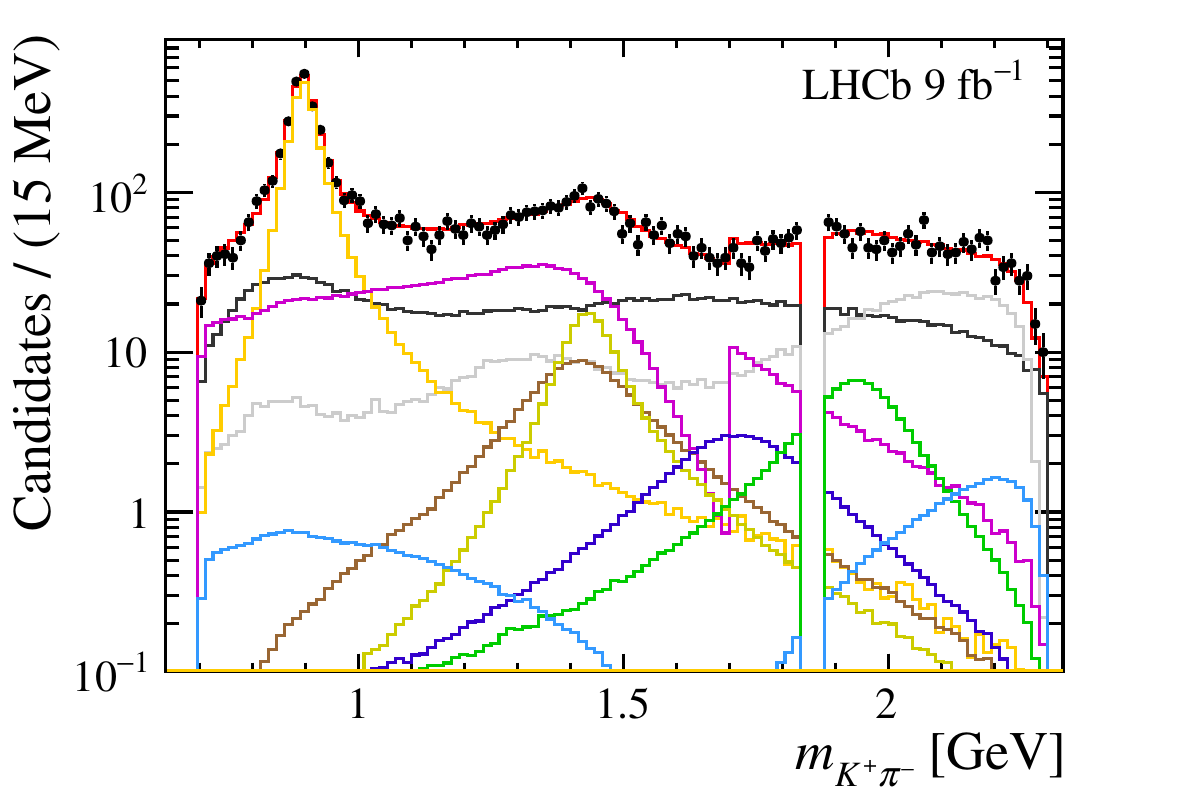}
\includegraphics[width=0.49\textwidth]{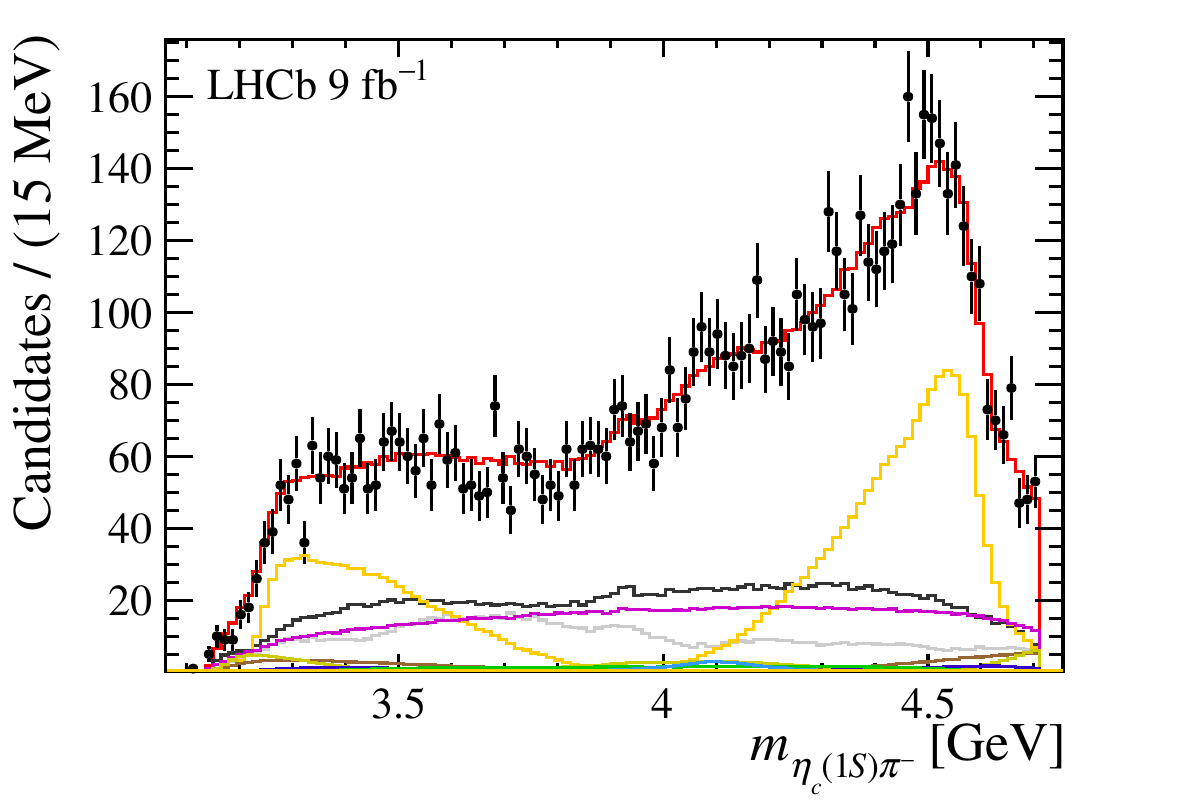}
\includegraphics[width=0.49\textwidth]{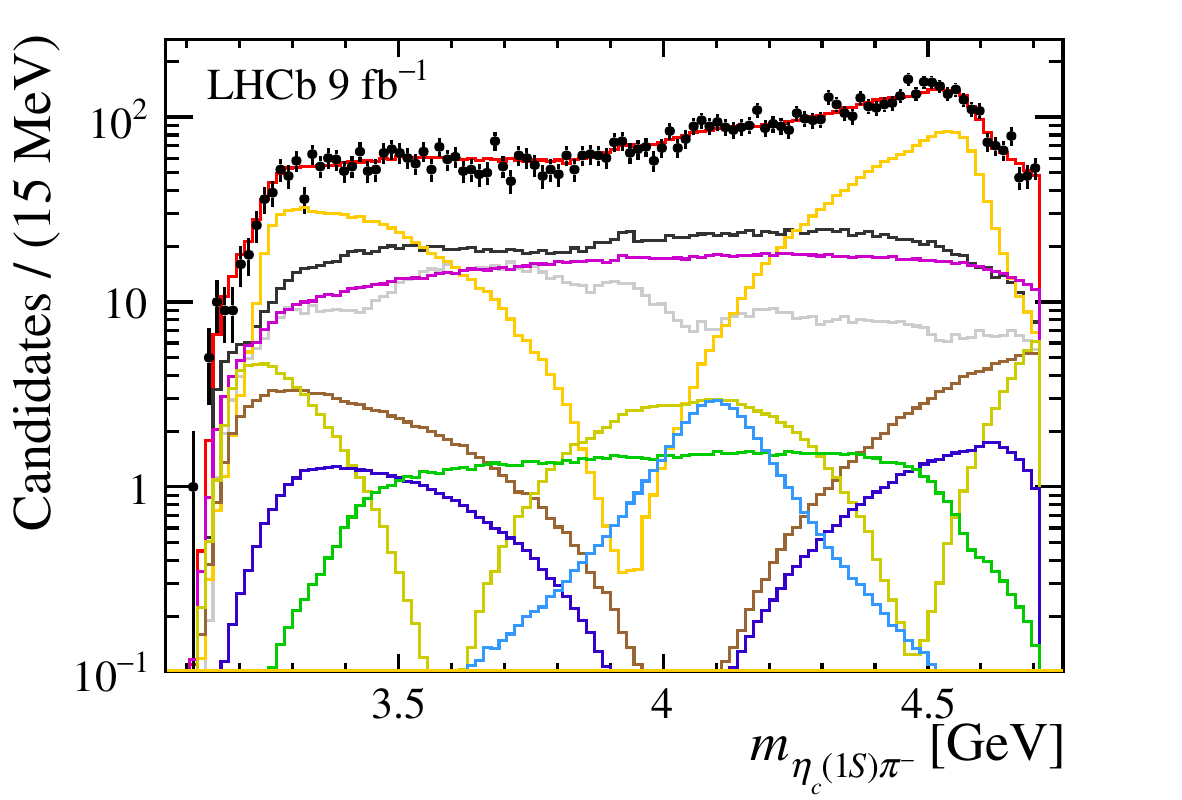}
\includegraphics[width=0.49\textwidth]{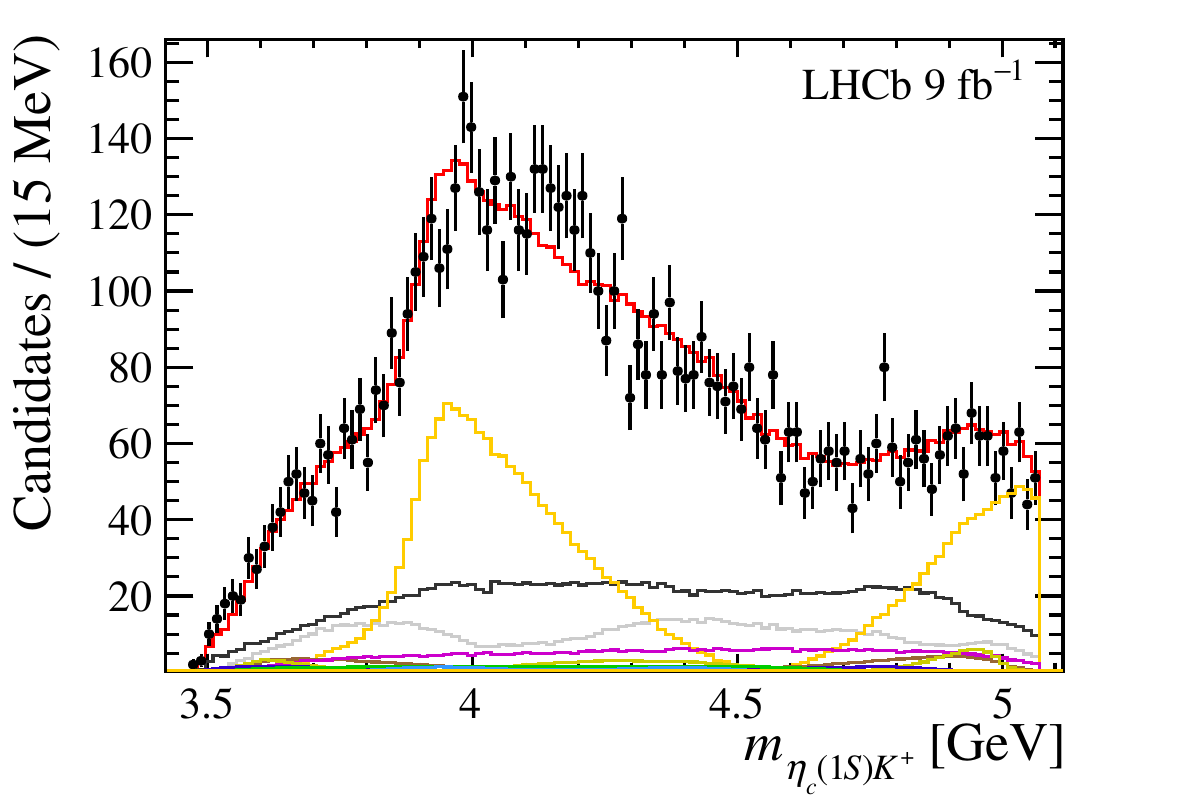}
\includegraphics[width=0.49\textwidth]{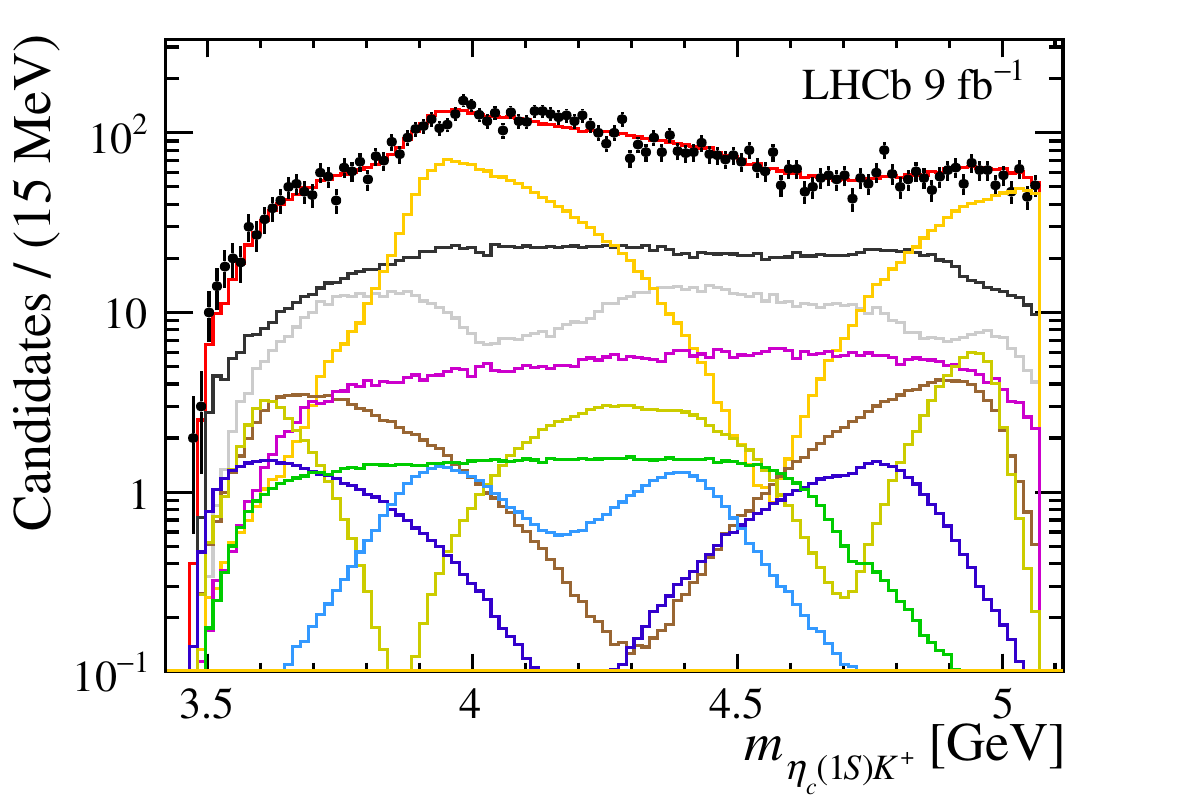}
\caption{Projections of the $\Bd\to\etac\Kp\pim$ data and DP fit using the extended model onto the (top)~$m_{\Kp \pim }$, (middle)~$m_{\etac \pim }$ and (bottom)~$m_{\etac \Kp }$ observables for (left) linear and (right) logarithmic vertical-axis scale. The veto of $\Bd  \to \proton \antiproton \Dzb$ decays is visible in the projections onto the~$m_{\Kp \pim }$ observable.}
\label{ZcPlots}
\end{figure}

A better description of the data, compared to the previous model, is sought by adding to the baseline model an amplitude corresponding to an exotic candidate, $T_{c\bar{c}}(4100)^- \to \etac  \pim $ which is described by a RBW, in addition to the \Kstarz resonances listed in Table~\ref{resonanceused}. In addition to the parameters that are free in the baseline model, the complex coefficients, and RBW mass and width of the exotic candidate are left free to vary.
The strategy followed for the LASS parameters is the same as described before ensuring faster convergence times when an exotic contribution is included in the model. The values found from the best fit for the LASS function and the exotic candidate are reported in Table~\ref{tabLASSZc}.
A likelihood-ratio test, based on the log-likelihood difference with respect to the baseline value, $\Delta(-2\ln\mathcal{L})$~\cite{stat}, is used to compare different amplitude models.
The exotic candidate is tested under two quantum-number hypotheses, repeating the DP fit for the $J^P=0^+$ and $1^-$ assignments.
The corresponding $\Delta(-2\ln\mathcal{L})$ values, are 16.2 and 29.0, respectively. Thus, the model providing the best description of data, referred to as the extended fit model, is obtained with the addition of an exotic candidate with $J^P=1^-$. 

\begin{table}[b]
\begin{center}
\caption{Parameters of the LASS function and the exotic $J^P=1^-$candidate resulting from the best fit using the extended model.}
\begin{tabular}{ l c c  }
\hline
 Parameter & Value & Lineshape\\
 \hline
 $r$ & \phantom{000}$4.6\pm\phantom{00}1.0\gev^{-1}$ & LASS\\
 $a$ &\phantom{000}$4.2\pm\phantom{00}0.7\gev^{-1}$  & LASS\\
 $m_{K^{*}_{0}(1430)}$&$1433\phantom{.0}\pm\phantom{0}19\phantom{.0}$\mev\phantom{0}  &  LASS\\
$\Gamma_{K^{*}_{0}(1430)}$&\phantom{0}$359\phantom{.0}\pm\phantom{0}37\phantom{.0}$\mev \phantom{0} & LASS\\
 $m_{T_{c\bar{c}}(4100)^{-}}$ &  $4106\phantom{.0}\pm\phantom{0}23\phantom{.0}$\mev\phantom{0}  & RBW\\
$\Gamma_{T_{c\bar{c}}(4100)^{-}}$ &$\phantom{0}506\phantom{.0}\pm166\phantom{.0}$\mev\phantom{0}  & RBW\\
 \hline
\end{tabular}
\label{tabLASSZc}
\end{center}
\end{table}
The values for the magnitude, phase and fit fraction of each contribution are reported in Table~\ref{Zcfit}.
Statistical uncertainties on the fit fractions are determined using large samples of pseudoexperiments generated from the fit results, thereby accounting for correlations between parameters. The mass projections of the extended model are shown in Fig.~\ref{ZcPlots}, for which the value of the \chisqndf is $ 401/305$.
The fit quality is further assessed comparing the unnormalised Legendre moments between background-subtracted data and the model.

\begin{table}[tb]
\begin{center}
\caption{Magnitudes, phases and fit fractions determined from the DP fit using the extended model. Uncertainties are statistical only.}
\begin{tabular}{  l c c c  }
\hline
 Amplitude & Magnitude & Phase & Fit fraction ($\%$)\\
 \hline
 $\Bd  \to \etac K^*(892)^0$ & 1 (fixed) & 0 (fixed) & $\phantom{0}49.1\pm1.3$\\ 
 $\Bd  \to \etac K^*(1410)^0$ & $0.28 \pm 0.03$ & $-0.04 \pm 0.15$ & $\phantom{00}3.9\pm0.9$\\ 
  $\Bd  \to \etac K_0^*(1430)^0$ & $0.73 \pm 0.04$ & $-0.18 \pm 0.05$ & $\phantom{0}26.2\pm4.1$\\ 
  $\Bd  \to \etac (\Kp\pim)_{\rm SVP}$ & $0.51 \pm 0.03$ & $+0.31 \pm 0.07$ & $\phantom{0}12.9\pm2.0$\\ 
 $\Bd  \to \etac K_2^*(1430)^0$ & $0.28 \pm 0.03$ & $-0.86 \pm 0.11$ & \phantom{00}$4.0\pm0.9$\\ 
 $\Bd  \to \etac K^*(1680)^0$ & $0.18 \pm 0.05$ & $+0.32 \pm 0.23$ & \phantom{00}$1.5\pm0.9$\\ 
 $\Bd  \to \etac K_0^*(1950)^0$ & $0.21 \pm 0.03$ & $-0.92\pm 0.35$ & \phantom{00}$2.2\pm0.7$\\  
 $\Bd  \to T_{c\bar{c}}(4100)^-\Kp $ & $0.15 \pm 0.03$ & $-3.14\pm 0.06$ & \phantom{00}$1.1\pm0.5$\\  
 \hline
 Sum of fit fractions & & & $100.9\pm3.5$\\
 \hline
\end{tabular}
\label{Zcfit}
\end{center}
\end{table}

The significance of the $T_{c\bar{c}}(4100)^{-}$ candidate is evaluated from the value of $\Delta(-2\ln\mathcal{L})$, assuming that this quantity follows a \chisq distribution
with degrees of freedom equal to twice the number of free parameters in its expression~\cite{LHCb-PAPER-2022-027,LHCb-PAPER-2022-040,LHCB-PAPER-2014-014}.
This assumption accounts for the look-elsewhere effect
due to the varying mass and width of the $T_{c\bar{c}}(4100)^{-}$ candidate. Its validity is verified through pseudoexperiments, which test the $\Delta(-2\ln\mathcal{L})$ distribution under the baseline hypothesis. The results confirm that this distribution is well described by a \chisq function with 8 degrees of freedom.
In the extended fit model, the statistical significance of the $T_{c\bar{c}}(4100)^{-}$  is found to be $3.6$ standard deviations~($\sigma$), excluding systematic uncertainties.
To distinguish between different $J^P$ assignments, fits are performed under alternative $J^P$ hypotheses. A lower limit on the significance of rejecting the $J^P=0^+$ hypothesis is determined from the change in the log-likelihood relative to the preferred hypothesis, assuming a $\chisq$ distribution with one degree of freedom. This assumption is similarly validated using pseudoexperiments, which confirm that the $\Delta(-2\ln\mathcal{L})$ distribution under the disfavoured
$J^P= 0^+$ hypothesis follows the expected \chisq behaviour. The $J^P=0^+$ hypothesis with respect to that of $J^P = 1^-$ is statistically disfavoured at $3.2\sigma$.
However, systematic effects must be taken into account to determine the significance of the $T_{c\bar{c}}(4100)^{-}$ contribution and the discrimination of its quantum numbers.

\subsection{Systematic uncertainties}
\label{Syst}
Systematic uncertainties on the fit fractions, amplitude, and phases are evaluated by measuring the difference between the baseline value and the value obtained from different studies as detailed below.

The systematic uncertainties associated with the background model are evaluated using alternative histograms, obtained by randomly varying the bin contents within their statistical uncertainties before spline interpolation. Approximately 1000 such histograms are generated for both the combinatorial and NR background components and used to construct an equal number of pseudoexperiments, where the signal contribution is generated according to the baseline model. Each pseudoexperiment is fitted with the baseline signal model and the corresponding modified background model, and the systematic uncertainty on the fit fractions is taken as the standard deviation of the resulting distribution of best-fit values, which is well described by a Gaussian.
The same pseudoexperiments are also used to estimate the significance of the $T_{c\bar{c}}(4100)^-$ state. Each is fitted twice: once with the baseline model and once with the extended model, both including the varied background maps. The resulting distribution of $\Delta(-2\ln\mathcal{L})$ values does not follow a \chisq distribution. A $p$-value is therefore determined from the fraction of fits yielding a $\Delta(-2\ln\mathcal{L})$ greater than that observed in data, corresponding to $1.45\%$, or a significance of $2.5\sigma$.

The strategy used to assess the systematic uncertainty associated to the background model is also adopted to account for the effects relating to the efficiency model.
Another systematic uncertainty is associated to the veto applied to the DP borders. The DP fit is repeated removing these vetoes.

The statistical uncertainties of signal and background yields are incorporated by applying constraints with Gaussian priors and repeating the fit to account for variations within these uncertainties.

A systematic uncertainty related to the potential fit bias is determined using pseudoexperiments. Samples are generated according to the baseline fit result and fitted again.
The associated error is calculated as the squared sum of the residuals and the standard deviation.

The kinematic variables used in the DP fit are computed from the \Bz-meson and final-state-particle four-momenta. 
 In order to associate a systematic uncertainty to the $\etac $ natural width,
the DP fits are repeated with kinematic variables that are recomputed assigning to the \etac mass the values $m_{\etac }\pm\Gamma_{\etac }/2$, where $m_{\etac }$ and $\Gamma_{\etac }$ are the mass and natural width of the $\etac $ meson,
respectively, obtained from the mass fits reported in Sec.~\ref{SignalDP}.

In the baseline model the \Kstarz resonances are modelled by RBW functions with their mass and width parameters fixed to known values~\cite{PDG2024}, and with fixed Blatt--Weisskopf barrier factor radii. The uncertainties related to this choice are determined by varying simultaneously and randomly the masses and widths within their uncertainties and, separately, by changing the radii between 3 and $5\gev^{-1}$.

The LASS model used to parametrise the low-mass $\Kp \pim $ S-wave in the baseline fit
is replaced with $K_0^{*}(1430)^{0}$ and $K_0^{*}(700)^{0}$ resonances parametrised by RBW functions, and a NR S-wave $\Kp \pim $ component parametrised with a uniform amplitude.

When including these sources of systematic uncertainties, summarised in Table~\ref{sigSyst}, the $T_{c\bar{c}}(4100)^-$ contribution is not found to be significant. Only the significances that deviate most from the nominal value are shown and the background parametrisation is found to be dominant. For this reason, systematic uncertainties associated to the mass and width of the $T_{c\bar{c}}(4100)^{-}$ candidate, and on the discrimination between quantum-number hypotheses, are not assigned. 

\begin{table}
\centering
    \caption{Significance of the extended model (with the $T_{c\bar{c}}(4100)^{-}$ candidate) with respect to the baseline model (without the $T_{c\bar{c}}(4100)^{-}$ candidate) for the main sources of systematic uncertainty.}
    \begin{tabular}{l c }
    \hline
    Source & Significance \\
    \hline
    Background parametrisation &  $2.5\sigma$\\
    Efficiency parametrisation & $4.1\sigma$\\
    DP-border veto & $3.2\sigma$\\
    Fixed yields & 3.2$\sigma$\\
    \etac natural width & 3.0$\sigma$\\
    Fixed \Kstarz lineshape parameters & $3.2\sigma$\\
    $\Kp \pim $ S-wave & $4.5\sigma$\\
    \hline
    \end{tabular}
    \label{sigSyst}
\end{table}

The baseline model is therefore used to compute the branching fractions of the \Kstarz contributions in order to characterise the $\Bd  \to \etac  \Kp  \pim $ decay channel. To this end, the extended model is also considered as an alternative to the baseline model, with associated systematic uncertainties. The addition of a resonance decaying to $\etac \Kp$ was initially considered, but it was consistently found to be less significant than the one decaying to $\etac \pim$. Therefore, it was not pursued further. The values for the fit fractions of the \Kstarz components and their interference fit fractions, together with their associated systematic uncertainties, are reported in Table~\ref{half-matrix_fit_default_LASS_separated}, while the magnitudes and phases are given in Table~\ref{magphasesyst}.

\begin{sidewaystable}
\begin{center}
\footnotesize
\caption{ \small{Symmetric matrix of the fit fractions and interference fit fractions~($\%$) resulting from the fit using the baseline model, where the quoted uncertainties are statistical and systematic, respectively. Entries marked with a dash correspond to instances where the modulus of the interference fit fraction is found to be less than $10^{-4}\%$.}}
\renewcommand{\arraystretch}{1.1}Jerryjerry1!
\resizebox{\textwidth}{!}{
\begin{tabular}{  l c c c c c c c }
\hline
 & $\etac K^*(892)^0$ &  $\etac K^*(1410)^0$ & $\etac K_0^*(1430)^0$ & $\etac (\Kp\pim)_{\rm SVP}$ & $\etac K_2^*(1430)^0$ & $\etac K^*(1680)^0$ &  $\etac K_0^*(1950)^0$ \\
\hline
$\etac K^*(892)^0$ & $49.3 \pm 1.2^{+3.8}_{-3.4}$ & $-2.1 \pm 1.8^{+3.1}_{-3.3}$ & --- & --- & --- & \phantom{+}$0.8 \pm 1.2^{+3.7}_{-4.5}$ & --- \\ 
$\etac K^*(1410)^0$ & & \phantom{+}$4.5 \pm 1.4^{+3.2}_{-3.1}$ & --- & --- & --- & $-3.2 \pm 1.5^{+6.5}_{-2.3}$ & --- \\ 
$\etac K_0^*(1430)^0$ & & & $30.8 \pm 4.7^{+5.5}_{-4.3}$ & $-10.8 \pm 1.3^{+1.9}_{-1.5}$ & --- & --- & $7.0 \pm 10.3^{+7.9}_{-8.6}$ \\ 
$\etac (\Kp\pim)_{\rm SVP}$ & & & & \phantom{+}$15.4 \pm 2.4^{+2.7}_{-2.2}$ & --- & --- & --- \\ 
$\etac K_2^*(1430)^0$ & & & & & $3.7 \pm 0.9^{+1.3}_{-2.0}$ & --- & --- \\ 
$\etac K^*(1680)^0$ & & & & & & \phantom{+}$2.1 \pm 0.9^{+1.3}_{-1.7}$ & --- \\ 
$\etac K_0^*(1950)^0$ & & & & & & & $2.4 \pm \phantom{0}0.5^{+2.7}_{-2.4}$ \\  
 \hline
\end{tabular}
\label{half-matrix_fit_default_LASS_separated}
}
\end{center}
\end{sidewaystable}

\begin{table}[tb]
\begin{center}
\caption{Magnitudes and  phases determined from the DP fit using the baseline model,  where the uncertainties are statistical and systematic, respectively.}
\begin{tabular}{  l c c }
\hline
 Amplitude & Magnitude & Phase\\
 \hline
 $\Bd  \to \etac K^*(892)^0$ & 1 (fixed) & 0 (fixed) \\ 
  $\Bd  \to \etac K^*(1410)^0$ & $0.30 \pm 0.04^{+0.06}_{-0.14}$ & $-0.14\pm0.13^{+0.57}_{-0.53}$\\ 
  $\Bd  \to \etac K_0^*(1430)^0$ & $0.79 \pm 0.04^{+0.17}_{-0.06}$ &$+2.94\pm0.05^{+0.09}_{-0.10}$\\ 
    $\Bd  \to \etac  (\Kp\pim)_{\rm SVP}$ & $0.56 \pm 0.03^{+0.15}_{-0.07}$ & $+2.81 \pm 0.07^{+0.07}_{-0.09}$\\ 
 $\Bd  \to \etac K_2^*(1430)^0$ & $0.28 \pm 0.03^{+0.03}_{-0.07}$ & $-0.73 \pm 0.10^{+0.19}_{-0.23}$\\ 
 $\Bd  \to \etac K^*(1680)^0$ & $0.21 \pm 0.05^{+0.05}_{-0.09}$ & $+0.21 \pm 0.22^{+0.59}_{-0.31}$\\ 
 $\Bd  \to \etac K_0^*(1950)^0$ & $0.22 \pm 0.03^{+0.09}_{-0.08}$ & $+1.53\pm 0.29^{+0.73}_{-0.55}$\\   
 \hline
\end{tabular}
\label{magphasesyst}
\end{center}
\end{table}

\section{Branching fraction measurement}
The measurement of the inclusive $\Bd \to{\etac \Kp \pim }$ branching fraction is performed relative to the normalisation channel $\Bd \to{\jpsi \Kp \pim }$.
The absolute branching fraction of the $\Bd  \to \etac  \Kp  \pim $ decay can then be calculated using the relation

\begin{equation}
    \mathcal{B}(\Bd  \to \etac  \Kp  \pim ) = \mathcal{B}(\Bd  \to  \jpsi \Kp  \pim )\frac{\mathcal{B}( \jpsi\to \proton\antiproton)}{\mathcal{B}(\etac \to \proton\antiproton)} \frac{N_{\etac }}{N_{ \jpsi}}\frac{\epsilon_{ \jpsi}}{\epsilon_{\etac }},
\end{equation}
where $\mathcal{B}(\Bd  \to  \jpsi \Kp  \pim ) = (1.15\pm0.05)\times 10^{-3}$, $\mathcal{B}( \jpsi \to \proton\antiproton) = (2.120\pm0.029)\times10^{-3}$ and $\mathcal{B}(\etac  \to \proton\antiproton) = (1.33\pm0.11)\times 10^{-3}$ are taken as external inputs~\cite{PDG2024}. The values $N_{\etac }$ and $N_{ \jpsi}$ are the respective yields of the $B^0 \to \eta_c K^{+} \pi^{-} $ and $ B^0 \to \jpsi K^{+} \pi^{-}$ contributions, as reported in Tables~\ref{Fit2Dres} and~\ref{Jpsiyield}, while $\epsilon_{ \jpsi}$ and $\epsilon_{\etac }$ are the selection efficiencies of the signal and normalisation channel calculated from simulation.

Selection efficiencies are determined from the $\Bd \to{\etac \Kp \pim }$ and $\Bd \to{ \jpsi \Kp \pim }$ simulated samples, both selected using the same criteria as applied to data. Weights are applied to the simulated candidates to correct for minor discrepancies with respect to data, based on track multiplicity, \Bz kinematic distributions, and the respective DP distributions. Discrepancies due to the trigger responses are investigated and found to be negligible. The ratio of efficiencies is given by $\epsilon_{ \jpsi}/\epsilon_{\etac }= 0.945\pm0.004$ for Run~1 and $\epsilon_{ \jpsi}/\epsilon_{\etac } = 0.948\pm0.003$ for Run~2, where the uncertainties are statistical.

Systematic uncertainties on the branching fraction measurement are evaluated by recomputing the signal and normalisation yields, along with their efficiency ratios, for each source of uncertainty. The difference of the resulting branching fraction from the baseline value is taken as the associated systematic uncertainty. The overall systematic uncertainty is calculated by combining all contributions in quadrature. The contribution arising from fixing the parameters of the Hypatia
PDFs used to parametrise the $\Bd $ and $ \jpsi$ components, as well as the Crystal Ball PDF for the $\etac $, is evaluated by repeating the fit with parameter values varied according to a multivariate normal distribution. To assess a systematic uncertainty for the model used to describe the mass resolution, the fits are repeated by replacing the Hypatia PDFs with Crystal Ball PDFs.
The convolution of the RBW with a Crystal Ball PDF is replaced by a Voigtian PDF, where the parameters are fixed based on simulation.
The  $m_{\proton\antiproton}$ distribution of the NR component, originally parametrised with an exponential function, is instead modelled with a linear function. The systematic uncertainty related to the choice of the mass ranges in the two-dimensional fit is evaluated by performing fits with alternative ranges. Table~\ref{TabSystBR} summarises the systematic uncertainties.
{ 
\begin{table}[tb]
\begin{center}
\caption{Systematic uncertainties on the $\Bd  \to \etac  \Kp  \pim$ branching fraction. Statistical uncertainties are also reported for comparison.}

\begin{tabular}{ l c c }
\hline
 &  \multicolumn{2}{c}{$\sigma_{\mathcal{B}(\Bd  \to \etac \Kp  \pim )} [10^{-4}]$} \\
 \hline
 Source & Run~1  & Run~2  \\
 \hline
Shape parameters & $0.05$& $0.05$\\
 Resolution and NR $\proton\antiproton$ model & $0.18$ & $0.22$\\
 Mass windows &$0.41$ & $0.15$\\
 \hline
 Total systematic uncertainty & 0.45 & 0.27\\
  \hline
  Statistical uncertainty & $0.53$ & $0.22$\\
  \hline

\end{tabular}
\label{TabSystBR}
\end{center}
\end{table}
}

The measured values of the branching fraction \mbox{$\mathcal{B}(\Bd  \to \etac \Kp  \pim )$} are  \mbox{$(6.10 \pm0.53\pm0.45\pm0.58)\times10^{-4}$} for Run~1 and $(5.77 \pm0.22\pm0.27\pm0.55)\times10^{-4}$ for Run~2.
This leads to the combined value
\begin{align*}
   \mathcal{B}(\Bd  \to \etac \Kp  \pim ) = (5.82\pm0.20\pm0.23\pm0.55)\times10^{-4},
\end{align*}
where the first uncertainty is statistical, the second systematic, and the third is due to the limited knowledge of the external branching fractions. The combined value is obtained as the weighted average of the Run~1 and Run~2 results. This result is in agreement with the world average~\cite{PDG2024} and with the previous LHCb result~\cite{LHCb-PAPER-2018-034}.
The product branching fractions for the intermediate states contributing to the $\Bd  \to \etac  \Kstarz(\to\Kp\pim)$ decays listed in Table~\ref{BR}, are determined by scaling the inclusive branching fraction $\mathcal{B}(\Bd  \to \etac \Kp  \pim )$ by the fit fractions of each intermediate state with the baseline model reported in Table~\ref{half-matrix_fit_default_LASS_separated}.
\begin{table}[tb]
\centering
    \caption{Product branching fractions for intermediate  $\Kstarz$ resonances decaying to $\Kp\pim$ contributing to the $\Bd  \to \etac  \Kstarz(\to\Kp\pim)$ decay, \mbox{$\BF(\Bz\to\etac \Kstarz) \times\BF(\Kstarz \to\Kp\pim)$}. The quoted uncertainties are statistical, from the inclusive branching fraction of $\Bz\to\etac\Kp\pim$, from the intermediate fit fraction and from the charmonium branching fractions of $\etac\to\proton\antiproton$ and $\jpsi\to\proton\antiproton$, respectively.}
    \begin{tabular}{l r }
    \hline
    Decay  & Branching fraction $[10^{-5}]$\\
    \hline
    \vspace{0.1cm}
    $\Bd  \to \etac  K^*(892)^0(\to \Kp \pim )$ & $28.69\pm1.21\pm1.13^{+2.21}_{-1.98}\pm2.71$\\
    \vspace{0.1cm}
    $\Bd  \to \etac K^*(1410)^0(\to \Kp \pim )$ &$2.62\pm0.82\pm0.10^{+1.86}_{-1.80}\pm0.25$\\
    \vspace{0.1cm}
    $\Bd  \to \etac K_0^*(1430)^0(\to \Kp \pim )$ &$17.92\pm2.80\pm0.71^{+3.26}_{-2.56}\pm1.69$\\
    \vspace{0.1cm}
    $\Bd  \to \etac (\Kp\pim)_{\rm SVP}$ &$8.96\pm1.42\pm0.35^{+1.63}_{-1.28}\pm0.84$\\
    \vspace{0.1cm}
    $\Bd  \to \etac K_2^*(1430)^0(\to \Kp \pim )$&$2.15\pm0.53\pm0.08^{+0.76}_{-1.16}\pm0.20$\\
    \vspace{0.1cm}
    $\Bd  \to \etac K^*(1680)^0(\to \Kp \pim )$&$1.22\pm0.52\pm0.05^{+0.76}_{-0.99}\pm0.12$\\
    \vspace{0.1cm}
    $\Bd  \to \etac K_0^*(1950)^0(\to \Kp \pim )$ &$1.40\pm0.29\pm0.06^{+1.57}_{-1.40}\pm0.13$\\

    \hline
    \end{tabular}
    \label{BR}
\end{table}

\section{Conclusion}

An amplitude analysis of the $\Bd  \to \etac(1S) \Kp  \pim $ decay channel is performed using the dataset collected by \lhcb in Run~1 and 2, corresponding to an integrated luminosity of 9\invfb. While a better description of the data is achieved when including a charged exotic charmonium-like $T_{c\bar{c}}(4100)^-$ candidate decaying to $\etac(1S) \pim $, the contribution is not significant when systematic uncertainties are taken into account. The results of this analysis are valid in the absence of additional suppressed intermediate resonances decaying to the $\Bd  \to \proton \antiproton \Kp  \pim$ final state, possibly interfering with the amplitudes involved in the $\Bd  \to (\etac(1S) \to \proton \antiproton) \Kp  \pim $ decay. Several checks, including the comparison between the Legendre moments computed from data and from the fit models, confirms that, with the current dataset, there is no sensitivity to such interference effects, if present.

A measurement of the inclusive $\Bd  \to \etac(1S)  \Kp  \pim $ branching fraction is also performed, as well as the branching fraction measurements of the intermediate \Kstarz resonances contributing to the decay. 
The results of this analysis, using a data sample size that is approximately doubled with respect to the previous study, supersede those obtained in Ref.~\cite{LHCb-PAPER-2018-034}.

%% file: acknowledgements.tex
\section*{Acknowledgements}
%
%
\noindent We express our gratitude to our colleagues in the CERN
accelerator departments for the excellent performance of the LHC. We
thank the technical and administrative staff at the LHCb
institutes.
We acknowledge support from CERN and from the national agencies:
ARC (Australia);
CAPES, CNPq, FAPERJ and FINEP (Brazil); 
MOST and NSFC (China); 
CNRS/IN2P3 (France); 
BMBF, DFG and MPG (Germany); 
INFN (Italy); 
NWO (Netherlands); 
MNiSW and NCN (Poland); 
MCID/IFA (Romania); 
MICIU and AEI (Spain);
SNSF and SER (Switzerland); 
NASU (Ukraine); 
STFC (United Kingdom); 
DOE NP and NSF (USA).
We acknowledge the computing resources that are provided by ARDC (Australia), 
CBPF (Brazil),
CERN, 
IHEP and LZU (China),
IN2P3 (France), 
KIT and DESY (Germany), 
INFN (Italy), 
SURF (Netherlands),
Polish WLCG (Poland),
IFIN-HH (Romania), 
PIC (Spain), CSCS (Switzerland), 
and GridPP (United Kingdom).
We are indebted to the communities behind the multiple open-source
software packages on which we depend.
Individual groups or members have received support from
Key Research Program of Frontier Sciences of CAS, CAS PIFI, CAS CCEPP, 
Fundamental Research Funds for the Central Universities,  and Sci.\ \& Tech.\ Program of Guangzhou (China);
Minciencias (Colombia);
EPLANET, Marie Sk\l{}odowska-Curie Actions, ERC and NextGenerationEU (European Union);
A*MIDEX, ANR, IPhU and Labex P2IO, and R\'{e}gion Auvergne-Rh\^{o}ne-Alpes (France);
Alexander-von-Humboldt Foundation (Germany);
ICSC (Italy); 
Severo Ochoa and Mar\'ia de Maeztu Units of Excellence, GVA, XuntaGal, GENCAT, InTalent-Inditex and Prog.~Atracci\'on Talento CM (Spain);
SRC (Sweden);
the Leverhulme Trust, the Royal Society and UKRI (United Kingdom).

%% file: Authorship_LHCb-PAPER-2025-027.tex
\centerline
{\large\bf LHCb collaboration}
\begin
{flushleft}
\small
R.~Aaij$^{38}$\lhcborcid{0000-0003-0533-1952},
A.S.W.~Abdelmotteleb$^{57}$\lhcborcid{0000-0001-7905-0542},
C.~Abellan~Beteta$^{51}$\lhcborcid{0009-0009-0869-6798},
F.~Abudin\'en$^{57}$\lhcborcid{0000-0002-6737-3528},
T.~Ackernley$^{61}$\lhcborcid{0000-0002-5951-3498},
A.A.~Adefisoye$^{69}$\lhcborcid{0000-0003-2448-1550},
B.~Adeva$^{47}$\lhcborcid{0000-0001-9756-3712},
M.~Adinolfi$^{55}$\lhcborcid{0000-0002-1326-1264},
P.~Adlarson$^{85}$\lhcborcid{0000-0001-6280-3851},
C.~Agapopoulou$^{14}$\lhcborcid{0000-0002-2368-0147}
% ...

{\footnotesize \it

% ...
$^{14}$Universit{\'e} Paris-Saclay, CNRS/IN2P3, IJCLab, Orsay, France\\
% ...
$^{38}$Nikhef National Institute for Subatomic Physics, Amsterdam, Netherlands\\
% ...
$^{47}$Instituto Galego de F{\'\i}sica de Altas Enerx{\'\i}as (IGFAE), Universidade de Santiago de Compostela, Santiago de Compostela, Spain\\
% ...
$^{51}$Physik-Institut, Universit{\"a}t Z{\"u}rich, Z{\"u}rich, Switzerland\\
% ...
$^{55}$H.H. Wills Physics Laboratory, University of Bristol, Bristol, United Kingdom\\
% ...
$^{57}$Department of Physics, University of Warwick, Coventry, United Kingdom\\
% ...
$^{60}$School of Physics and Astronomy, University of Glasgow, Glasgow, United Kingdom\\
$^{61}$Oliver Lodge Laboratory, University of Liverpool, Liverpool, United Kingdom\\
% ...
$^{69}$Syracuse University, Syracuse, NY, United States\\
% ...
$^{85}$Department of Physics and Astronomy, Uppsala University, Uppsala, Sweden, associated to $^{60}$\\
% ...
}
\end{flushleft}